\documentclass[reprint,showpacs,preprintnumbers,amsmath,amssymb, citeautoscript,aip,superscriptaddress]{revtex4-1}

\usepackage{graphicx}
\usepackage{dcolumn}
\usepackage{bm}
\usepackage{here}
\usepackage{hyperref}

\usepackage{caption}
\usepackage{subcaption}
\usepackage{xcolor}

\begin{document}


\title{Principles of Low Dissipation Computing from a Stochastic Circuit Model}
\author{Chloe Ya Gao}
\affiliation{Department of Chemistry, University of California, Berkeley, California \looseness=-1}
\author{David T. Limmer}
\email{dlimmer@berkeley.edu}
\affiliation{Department of Chemistry, University of California, Berkeley, California \looseness=-1}
\affiliation{Kavli Energy NanoScience Institute, Berkeley, California \looseness=-1}
\affiliation{Materials Science Division, Lawrence Berkeley National Laboratory, Berkeley, California \looseness=-1}
\affiliation{Chemical Science Division, Lawrence Berkeley National Laboratory, Berkeley, California \looseness=-1}

\date{\today}

\begin{abstract}
We introduce a thermodynamically consistent, minimal stochastic model for complementary logic gates built with field-effect transistors. We characterize the performance of such gates with tools from information theory and study the interplay between accuracy, speed, and dissipation of computations. With a few universal building blocks, such as the NOT and NAND gates, we are able to model arbitrary combinatorial and sequential logic circuits, which are modularized to implement computing tasks. We find generically that high accuracy can be achieved provided sufficient energy consumption and time to perform the computation. However, for low-energy computing, accuracy and speed are coupled in a way that depends on the device architecture and task. Our work bridges the gap between the engineering of low dissipation digital devices and theoretical developments in stochastic thermodynamics, and provides a platform to study design principles for low dissipation digital devices.
\end{abstract}

\maketitle

\section{Introduction}
The last decade has seen an exponential growth in energy consumption associated with information, communications, and computing technologies. Such resource demands are not sustainable, and thus there is a need to design devices with reduced energetic costs. While the problem of computing efficiency dates back to Landauer \cite{landauer1990maxwell,bennett2003notes}, with modern developments in stochastic thermodynamics, this problem is actively being revisited \cite{parrondo2015thermodynamics,wolpert2019stochastic}. The main goal of this paper is to bridge the gap between developments in nonequilibrium statistical physics and  circuit engineering by proposing a model for stochastic logic circuits that is thermodynamically consistent, and thus amenable to physical analysis and constraints, but simple enough to be extendable to complex computing tasks. By treating thermal fluctuations in electron transport explicitly at a mesoscopic scale, our model reproduces the behavior of a robust circuit in the low noise limit, but describes errors accurately away from this limit. With this model we explore the consequences of carrying out computations at low thermodynamic costs and finite time, and provide design principles for low dissipation computing devices.   

State-of-the-art semiconductor devices are typically built from metal-oxide-semiconductor field-effect transistors on the scale of a few nanometers, enabling billions of transistors to be packed on a single chip. In order to mitigate heating and large energy consumption burdens, it would be advantageous to operate such small devices with small bias voltages; however, as biases approach thermal scales, fluctuations increase, which necessitates a careful treatment of thermal noise \cite{johnson1928thermal,nyquist1928thermal}. The conventional treatment of thermal noise is largely phenomenological and involves either a  correction to the power spectral density \cite{van1970noise}, or transformation of the internal noise into external independent sources \cite{heinen1991unified,rizzoli1988state}. Such models are typically valid only near equilibrium where the fluctuation-dissipation theorem can be invoked to constrain their functional form \cite{maes2014second}, whereas higher-order correlations are needed in general to determine the full response \cite{speck2006restoring,gao2019nonlinear,lesnicki2020field}. While these models can provide insight into how thermal noise may put a physical limit on the density of transistors~\cite{kish2002end}, their validity in non-linear electrical networks operating far from equilibrium is uncertain.

Stochastic thermodynamics provides a theoretical way to move beyond an equilibrium description of thermal noise and its impact on information processing~\cite{seifert2012stochastic}. 
While information theory provides limits on the accuracy of typical communication \cite{shannon1948mathematical,cover1999elements}, stochastic thermodynamics provides generalized fluctuation-dissipation relationships, and places limits on the work required to implement a physical process in finite time and the spectrum of its fluctuations \cite{gallavotti1995dynamical,jarzynski1997nonequilibrium,crooks1999entropy,jarzynski2011equalities,neri2020second}. The link between information theory and stochastic thermodynamics has generated a wealth of expressions relating precision, speed, and dissipation, including the thermodynamic uncertainty relationships, speed limits, and fluctuation theorems. For example, dissipation bounds the rate at which a system transforms between different states \cite{roldan2015decision,shiraishi2018speed,ito2018stochastic,ito2020stochastic,falasco2020dissipation,kuznets2021dissipation,neri2021dissipation}. Dissipation also provides an upper bound for the precision of a current \cite{gingrich2016dissipation,gingrich2017inferring,horowitz2017proof}. A universal tradeoff between power, precision, and speed has been proposed for communication systems as well \cite{lahiri2016universal}. These theoretical results have  found application in many biological processes that natively operate near thermal energy scales~\cite{murugan2012speed,rao2015thermodynamics,barato2015thermodynamic,banerjee2017elucidating,pineros2020kinetic,wang2020price}. Placed in the context of artificial computing, these relationships have shed light on fundamental constraints on the design of computing devices to minimize thermodynamic costs \cite{plenio2001physics,still2012thermodynamics,sagawa2013information,parrondo2015thermodynamics,wolpert2019stochastic,wolpert2020thermodynamics}.

While such theoretical results are general, to apply them to the problem of computing design requires a realistic physical representation of information processing, such as bit storage, measurement, and erasure. Some success has been made with nonlinear single-electron devices and Coulomb blockade systems \cite{devoret1990effect,wasshuber2001computational,bagrets2003full}, where the logical states are represented by the presence of a few electrons. More recently, thermodynamically consistent stochastic models have been proposed for transistors and non-linear electronic circuits using either the continuous or discrete degrees of freedom \cite{gu2019microreversibility,gu2020counting,freitas2020stochastic}. For example, two-terminal devices, such as tunnel junctions, diodes, and metal-oxide-semiconductor (MOS) transistors, have been modeled as bi-directional Poisson processes embedded in a Markovian graph representing electron transfer \cite{freitas2020stochastic}. While such models can reproduce nonlinear current behaviors and noise characteristics, the nonlinearity has to be encoded by parametrizing the voltage dependence of the forward and backward rates. In this paper, we adopt a different approach where single logic gates are described by a tunnel junction model on the mesoscopic scale, combined with a capacitive circuit model for the charging and manipulation of the device. In this case, nonlinearity emerges from many interacting gates. Such an approach is able to describe electron transport processes consistent with the fluctuation theorems \cite{esposito2007fluctuation,nakamura2011fluctuation}, but also consistent with the complementary metal-oxide-semiconductor (CMOS) circuit platform used widely in modern computing devices. Therefore, it provides an ideal platform to study circuit behaviors with the tool of stochastic thermodynamics. 

In what follows, we demonstrate principles for low dissipation computing by constructing a stochastic model for logic circuits from a bottom-up approach. By working with elementary linear components, we can build nonlinear circuits that are thermodynamically consistent. 
 We first introduce a model for single gates, including the NOT gate and the NAND gate, and discuss their physical properties. We then study the collective behaviors of these basic components, including spatial correlations within combinational circuits, and temporal correlations within sequential circuits, where the emphasis will be on circuit design principles. The logic circuits are finally modularized and scaled up to a computing device to illustrate how multiple components are synchronized to complete a computing task. Throughout, the thermodynamically consistent model enables a description of errors and dissipation.

\section{Model for Single Gates}

Modern CMOS circuits implement logic functions 
by integrating two different types of transistors differentiated by their major charge carriers, so-called N-type and P-type transistors. Here we choose a mesoscopic tunnel junction model to describe electron transport in a single gate \cite{datta1997electronic}. The transistors are modeled by two single-electron levels of energy $\epsilon_i$ with $i={\mathrm{N},\mathrm{P}}$ for the N-type and P-type transistors. The electrodes are modeled by electron reservoirs with chemical potential $\mu_j$ with $j=s,d,g$ denoting the source, drain, and gate, respectively. Electron transfer among them is described by a Markovian master equation, parametrized by transition rates $k_{ji}$ that describe the exchange rate of an electron from site $i$ to $j$. The transition rates are chosen to satisfy a local detailed balance condition, and thus guaranteeing thermodynamical consistency,
\begin{equation}
\frac{k_{ji}}{k_{ij}}=e^{-\beta (E_j-E_i)}\\
\end{equation}
where $\beta=1/k_\mathrm{B}T$, $k_\mathrm{B}$ is the Boltzmann constant, and $T$ the temperature of the device. The energy is described by either the band energy for an electron in the transistor, $\epsilon_i$, or a chemical potential, $\mu_j$, for an electron in an electrode. The condition of local detailed balance is a prerequisite for the application of stochastic thermodynamics, as it ensures a correct description of dissipation away from equilibrium, and relaxation to a Boltzmann distribution at equilibrium. While local detailed balance models each microscopic transition as being thermally mediated, emergent nonlinear behaviors resulting from collections of transitions can take the system arbitrarily far from equilibrium~\cite{wachtel2018thermodynamically}.

The energy levels of the transistors are controlled by an input voltage denoted $V_{\textrm{in}}$. In the case of a field-effect transistor, $V_{\textrm{in}}$ refers to the gate voltage that switches the transistor on and off.
In the limit of high gate capacitance, $V_{\textrm{in}}$ changes the energy levels of the transistors approximately linearly \cite{datta1997electronic}
\begin{equation}
\epsilon_{\textrm{P}}=\epsilon_{\textrm{P}}^{0} + qV_{\textrm{in}},\,\quad \epsilon_{\textrm{N}}=\epsilon_{\textrm{N}}^{0} - q V_{\textrm{in}},
\label{eq:epsilon}
\end{equation}
where $\epsilon_{{i}=\mathrm{N,P}}^{0}$ are reference energies and $q$ is the unit of electric charge. The sign of the slope differentiates the N and P-type transistors with different charge carriers. In our model, a voltage also uniquely determines the energetics of the electrodes by modulating their chemical potentials, $\mu_j=-q V_j$. Throughout, we will differentiate between two different types of electrodes. The first type, including the source and drain electrodes, is kept at fixed potentials, $V_s$ and $V_d$, respectively.  The second type, the gate electrode, satisfies a capacitive charging model with a fluctuating voltage $V_g$ for reading out a gate. This is justified by the fact that in CMOS circuits, the output of a single gate is usually  used as the input of another gate, in which case the two are connected through a capacitor.  The dynamics of $V_g$ is described by the equation of motion
\begin{equation}
C_g \frac{dV_g}{dt}=-J_g(t),
\label{eq:capacitor}
\end{equation}
where $C_g$ is the capacitance and $J_g$ is the electron current flowing into the electrode from the transistors. The constant capacitance implies a quadratic energy for charging the electrode, $E = C_g V_g^2/2$.

We adopt a semi-classical ballistic transport model for the rate of transfer of an electron from an electrode into or out of a transistor \cite{datta1990simple,harbola2006quantum,leijnse2008kinetic}. Such a description is valid in the weak coupling limit between a transistor and an electrode relative to the thermal energy, and for transistors that are small in scale relative to the mean free path of the electron. We restrict our analysis to single energy level transistors, for which the corresponding transition rates between transistor $i$ and electrode $j$ are
\begin{equation}
k_{ij}=\Gamma f_j(\epsilon_{i}), \, \quad k_{j{i}}=\Gamma [1-f_j(\epsilon_{i})],
\end{equation}
where $f_j(x)=[e^{\beta(x-\mu_j)}+1]^{-1}$ is the Fermi distribution. 
The prefactor $\Gamma$ is related to contact resistances and is chosen so that the timescale of electron transitions is longer than the timescale of thermal fluctuation, and thus the broadening of energy levels due to the coupling is smaller than thermal fluctuations. In making these assumptions to simplify our model, we have neglected effects such as scattering within the transistor, delocalization between the electrode and the transistor, and electron correlations, each of which can be incorporated into our model as long as thermodynamical consistency is retained.

Since we will be considering energy scales on the order of thermal fluctuations at the room temperature, we use $V_T=k_BT/q\approx 26 \mathrm{m}e\mathrm{V}$ and $\beta\hbar\approx 25$fs as our units of voltage and time, where $\hbar$ is Planck's constant. The voltage signal-to-noise ratio $V_d/V_T$ in our model will be on the order of 10, which is the prerequisite of low dissipation in the computing process since the two are closely related. While this ratio is much lower than the current technology, and requires delicate operation of the device, it can be experimentally achieved by designs such as the single-electron box \cite{koski2013distribution,koski2014experimental}. 
We reference potentials relative to the source voltage so that $V_s=0$, and take $\epsilon_\mathrm{P}^0 =0$ and $\epsilon_\mathrm{N}^0 =1.5 q V_d$ so that there exists only one independent energy parameter $V_d$. The transition rate constant is chosen as $\beta\hbar \Gamma=0.2$ to ensure the weak coupling assumption is valid~\cite{breuer2002theory}. To study the dynamics of the gates, we use both the exact steady-state solution of master equation when possible, and Gillespie simulations \cite{gillespie1976general} to sample individual trajectories.  We set  $C_g=200q/V_T$ in order to separate the timescales of capacitor charging from individual electron transfer events, simplifying the Gillespie simulations. Details of the numerical methods and the justification of the parameters can be found in Section I of the Supplementary Information ($\textit{SI}$).

\begin{figure}[t]
\centering
\includegraphics[width=6.2cm]{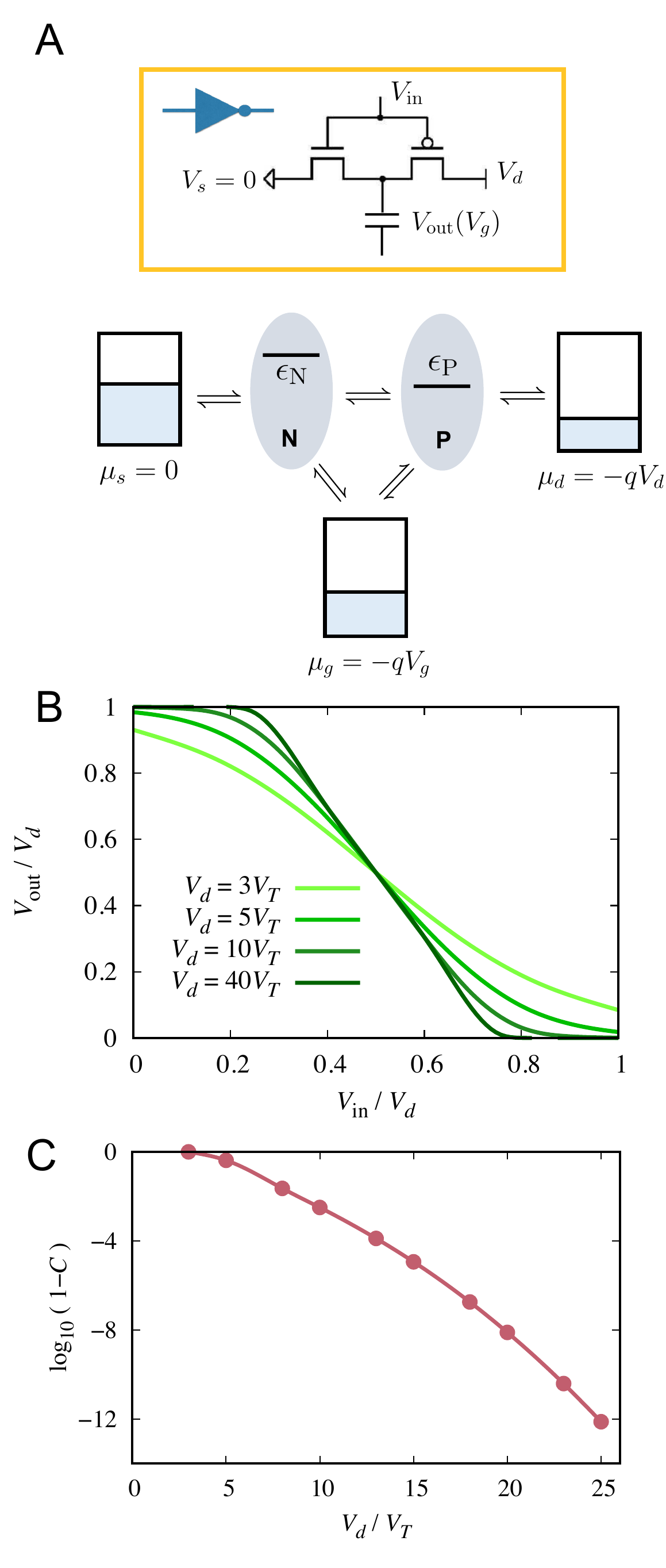}
\caption{Performance of a single NOT gate. (\textit{A}) Circuit diagram (above) and kinetic diagram (below) of an NOT gate, which is composed of a N-type (left) and a P-type transistor (right). 
(\textit{B}) Voltage transfer curve of a NOT gate. 
(\textit{C}) Channel capacity improves with increasing drain voltage $V_d$. 
}
\label{fig:NOT}
\end{figure}

\subsection{NOT gate}
The NOT gate, also known as the inverter, takes a single binary input $X$, and generates its complement as the output $Y$. The circuit diagram of the NOT gate, composed of two transistors, is shown in Fig. \ref{fig:NOT}$A$. The N-type transistor is connected to a lower source voltage $V_s=0$ on its left, while the P-type transistor is connected to a higher drain voltage $V_d$ to its right. Both transistors are controlled by an input voltage $V_{\textrm{in}}$ as in Eq.~\ref{eq:epsilon}, which is treated as fixed in a single gate, while the output voltage $V_{\textrm{out}}$ is measured between the two transistors from the capacitor voltage $V_g$, which evolves according to Eq.~\ref{eq:capacitor}. 
The kinetic diagram for our Markovian model is also shown in Fig. \ref{fig:NOT}$A$. Electrons can move ballistically between adjacent sites in the kinetic diagram according to a master equation, the details of which can be found in the $\textit{SI}$ (Eq. S4).

A NOT gate is typically characterized by its voltage transfer curve (VTC), shown in Fig. \ref{fig:NOT}$B$. The VTC reports on the average $V_{\textrm{out}}$ in response to $V_{\textrm{in}}$ in the long time limit. Generically, we find increasing $V_{\textrm{in}}$ results in a decrease in $V_{\textrm{out}}$ in agreement with the expected response of an inverter. However, its behavior is dependent on the scale of the thermal noise relative to $V_d$. The limiting values of $V_{\textrm{out}}$ approach $0$ and $V_d$ for $V_\mathrm{in}=V_d$ and 0, respectively, and sharpens between these limits with increasing $V_d$. Both features result from tuning the band energies of the two transistors in or out of resonance with their respective electrodes, as the transistor band energies depend on $V_d$ through Eq.~\ref{eq:epsilon}. Increasing $V_d$ with $V_{\textrm{in}}=0$ or $V_d$, increasingly suppresses current into the gate capacitor from $V_d$ or $V_s$. In the limit that current flows from only one electrode with fixed voltage, the gate electrode would reach an equilibrium state with that same voltage. The approach to this limiting behavior is exponential, for example, for increasing $V_d \gg V_T$ and $V_\mathrm{in}=0$, $|V_{\textrm{out}}-V_d | \sim \exp[-V_d /2 V_T]$.
The VTC is also symmetric around $V_{\textrm{in}}=V_{\textrm{out}}=V_d/2$, under which condition the difference between the energy level of the transistors and its connecting reservoirs is roughly the same for the N-type and P-type transistors.

\subsubsection{Performance as a computing unit}
When used as a computing unit, our first concern is whether our model generates the correct output with high probability. We define a perfect gate or device as one that generates a deterministic output according to the truth table, e.g., $Y$ should be the complement of $X$ for a perfect NOT gate. However, in the presence of noise, the deterministic output becomes stochastic and subject to finite error rates. As can be anticipated from the behavior of the VTC, in the limit of high $V_d/V_T$, or the low-noise limit, the performance of our model approaches that of a perfect NOT gate, whereas the behavior is nontrivial at smaller $V_d$.

The input and output signals are given as voltages in this model, so we map them to binaries by 
\begin{equation}
X=\left\{\begin{matrix}
0, & V_{\textrm{in}}=0\\
1, & V_{\textrm{in}}=V_d
\end{matrix}\right.\hspace{5mm}
Y=\left\{\begin{matrix}
0, & V_{\textrm{out}}\leq \alpha V_d\\
1, & V_{\textrm{out}}\geq (1-\alpha) V_d\\
\varnothing, & \textrm{otherwise}
\end{matrix}\right.
\label{eq:digital}
\end{equation}
where $\varnothing$ represents an invalid result that cannot be designated and $\alpha$ represents an error tolerance with $0< \alpha \ll 1$. We choose $\alpha=0.02$ so that the resultant error is below $10^{-10}$ for $V_d=40 V_T$ as comparable to current technologies, but our qualitative results are insensitive to this choice. 

To characterize the accuracy of the gate, we define the error rate $\xi$ as the probability of observing an output different from the perfect gate in a single shot. In the case of $X=0$, the error rate can be calculated from the empirical distribution of $V_\textrm{out}$ in steady state, as $\xi(X=0)=p[V_\textrm{out}<(1-\alpha)V_d | V_{\textrm{in}}=0]= 0.36$. A comprehensive characterization of the accuracy that takes into account the error rate for both cases of $X=0/1$ is the channel capacity 
\begin{equation}
C=\max_{p(X)} I(X;Y),
\end{equation}
which is the highest information rate that can be achieved with arbitrarily small error \cite{shannon1948mathematical}.
We compute $C$ numerically from the mutual information $I(X;Y)$ between the input and output at steady state as a function of $V_d$ (see details in \textit{SI} Section II), as shown in Fig.~\ref{fig:NOT}C. For a binary channel, the capacity is between 0 and 1, with 1 corresponding to a perfect gate. Here the capacity is computed to be $C=0.60$ for a channel operated at $V_d=5V_T$, given the slight difference between the error rate for $X=0$ or 1. 
While from the VTC the mean $V_\mathrm{out}$ is influenced by both the source and drain electrode for finite $V_d$, we find its distribution to be Gaussian with variance $1/(\beta C_g)$ within the steady state (Fig. S1). This is expected from a Boltzmann distribution, reflecting a proximity to equilibrium despite the presence of persistent currents. To reach a higher capacity, we need the average output $V_{\textrm{out}}$ to approach the limits 0 or $V_d$. This can be achieved by operating at a higher $V_d$ so that the leakage current flowing through the higher energy level transistor is even smaller. Given the Gaussian statistics, asymptotically for large $V_d$ the error scales as $\xi\sim \exp[-\beta C_g \alpha^2 V_d^2 /2] \sqrt{2 \pi /\beta C_g}/\alpha V_d$ and the channel capacity scales as $C\sim 1- \xi (1-\log_2 \xi)$, consistent with Fig.~\ref{fig:NOT}C. 

\begin{figure}[t]
\centering
\includegraphics[width=0.85\linewidth]{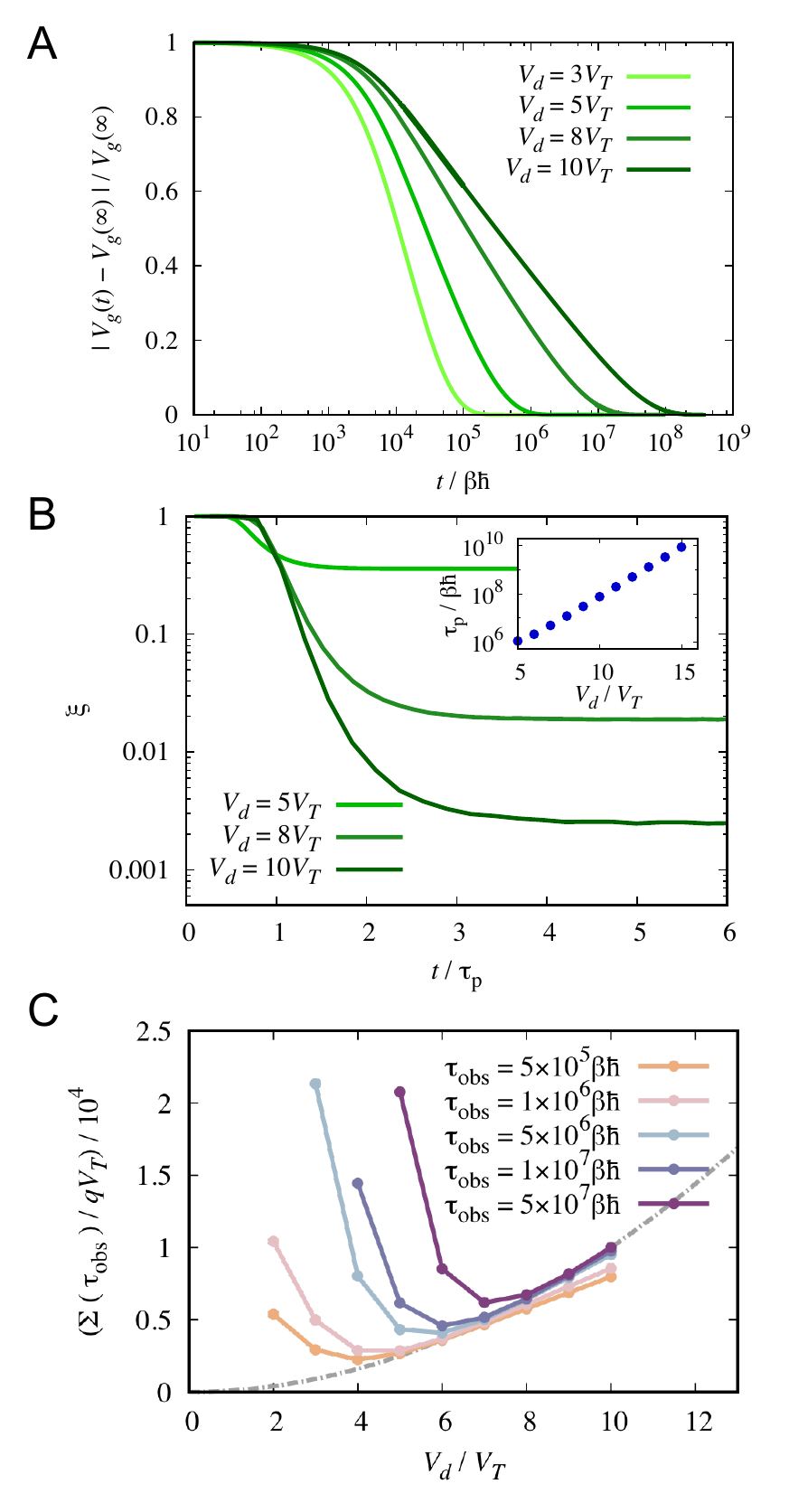}
\caption{Trade-off among accuracy, speed, and dissipation for a single NOT gate. (\textit{A}) Relaxation towards the steady state for a NOT gate initialized with $V_g=0V_T$, and $X=0$. (\textit{B}) The decay of the error rate with time, scaled by propagation delay $\tau_p$ . (\textit{Inset}) Propagation delay as a function of $V_d$.  (\textit{C}) The heat dissipation is a non-monotonic function of $V_d$ for finite observation time $\tau_{\textrm{obs}}$. The gray dashed line is the reversible limit $C_gV_d^2/2$.}
\label{fig:Vd}
\end{figure}

\subsubsection{Trade-off among accuracy, speed and dissipation}
While the accuracy of the gate improves dramatically for $V_{d} \gg V_T$, its performance is compromised by significantly increasing costs in computation time and energy consumption. Upon receiving a distinct input signal, the gate requires time to charge or discharge the capacitor to reach a steady state output signal. The average relaxation to steady state is shown in Fig. \ref{fig:Vd}$A$ for an initially discharged capacitor with input $X=0$. The relaxation is monotonic and nearly exponential but with characteristic decay time that depends on $V_d$.  Under this initial condition and input voltage, $\epsilon_{\textrm{N}} \gg \mu_s$, so that few electrons can flow between the source and the capacitor. The lower energy level $\epsilon_{\textrm{P}}$ facilitates electrons to transfer from the capacitor to the drain following the concentration gradient, gradually building up a higher voltage. 

We define the time it takes for $V_\mathrm{out}$ to reach $(1-\alpha)V_d$, the threshold voltage for $Y=1$, as the propagation delay time $\tau_p$. While the threshold voltage increases linearly with $V_d$, the average propagation delay $\tau_p$ grows exponentially. The propagation delay time, $\tau_p$ follows an inverse Gaussian distribution\cite{crooks2019field} with a long exponential tail (Fig. S1). Note that $\tau_p$ coincides with the time required for the error rate to decay below 0.5. Figure~\ref{fig:Vd}$B$ shows the decay of the error rate with time for $V_d=5,8,10V_T$, scaled by the propagation delay $\tau_p$ for each $V_d$.  As the distribution of $V_\mathrm{out}$ remains Gaussian, the time dependence of the error reflects the charging of the gate capacitor, and specifically follows the evolution of the mean $V_\mathrm{out}$. While we consider the single-shot error, the exponential scaling of $\tau_p$ with $V_d$ implies that associating an error rate with a time-averaged measurement of $V_\mathrm{out}$ would yield a non-monotonic relationship between the waiting time to reach a set error threshold and $V_d$. For intermediate $V_d$, the slower decorrelation time will cause waiting times to increase with $V_d$, while for large $V_d$ the suppressed fluctuations will dominate and decrease waiting times.

When the gate is used repeatedly to process a sequence of inputs $\textbf{X}=\{X_1,X_2,\cdots,X_{N}\}$, there is no need to re-initialize the gate after each computation, and the residual charge on the capacitor may help reduce the computational cost. We call this a memory effect, which introduces temporal correlation between consecutive data transmission processes. For such an information channel with memory, the accuracy can be characterized with the average information rate per data, which is a generalization of the channel capacity \cite{verdu1994general}, and the detail of which can be found in the $\textit{SI}$ (Section III). Using this metric, we find the memory effect plays a significant role at intermediate $\tau_{\textrm{obs}}$ enhancing the robustness of transmission by up to 30 percent (Fig. S2). For times much longer than $\tau_p$, the memory effect wears off and the information rate is set by the channel capacity. 

The energy consumption for a gate can be quantified with the heat dissipated to the environment.  From stochastic thermodynamics, the heat dissipation of the NOT gate during a long observation time $\tau_{\textrm{obs}}$ can be computed by the product of electron current and its conjugate affinity from two separate pathways \cite{gu2020counting}
\begin{equation}
\Sigma(\tau_{\textrm{obs}})=\int_0^{\tau_{\textrm{obs}}} dt\,\, J_{s\rightarrow \textrm{N}}(\mu_s-\mu_g)+J_{d\rightarrow\textrm{P}}(\mu_d-\mu_g),
\label{eq:entropy}
\end{equation}
where $J_{s\rightarrow \textrm{N}}$ is the electron current flowing from the source to the N-type transistor, and $J_{d\rightarrow\textrm{P}}$ is the current from the drain to the P-type transistor (Eq. S5). In the process described in Fig. \ref{fig:Vd}$A$, the pathway through the N-type transistor is essentially blocked due to the high-energy level of $\epsilon_{\textrm{N}}$, so the main contribution in Eq.~\ref{eq:entropy} is the second term in the sum. This second term has a similar form as the work required to quasi-statically charge the capacitor from $V_g=0$ to $V_g\approx V_d$, and thus is close to $C_gV_d^2/2$. This initial charging process is the dominant contribution to the heat dissipation over short times, and represents the reversible limit of the NOT gate (Fig. S1). Once the system reaches the steady state, there is still a steady entropy production coming from the leakage currents through both pathways, but the entropy production rate within the steady state is much smaller and decreases exponentially with $V_d$ (Fig. S3). This is because the output voltage $V_{\textrm{out}}$ is very close to $V_d$, leaving the affinity across the drain and the output nearly zero. Further, the corresponding leakage current from the source to the output is  small due to the high-energy level $\epsilon_{\textrm{N}}$.  The contributions to $\Sigma(\tau_{\textrm{obs}})$ from $V_d$  implies that for each observation time $\tau_{\textrm{obs}}$, there exists an optimal $V_d$ that minimizes $\Sigma(\tau_{\textrm{obs}})$, as confirmed in Fig. \ref{fig:Vd}$C$. The minimum $V_d$ shifts to the right with increasing time as at higher $V_d$ a larger  contribution from the steady-state flux counterbalances the higher heat dissipation during charging.

\begin{figure}[t]
\centering
\includegraphics[width=0.8\linewidth]{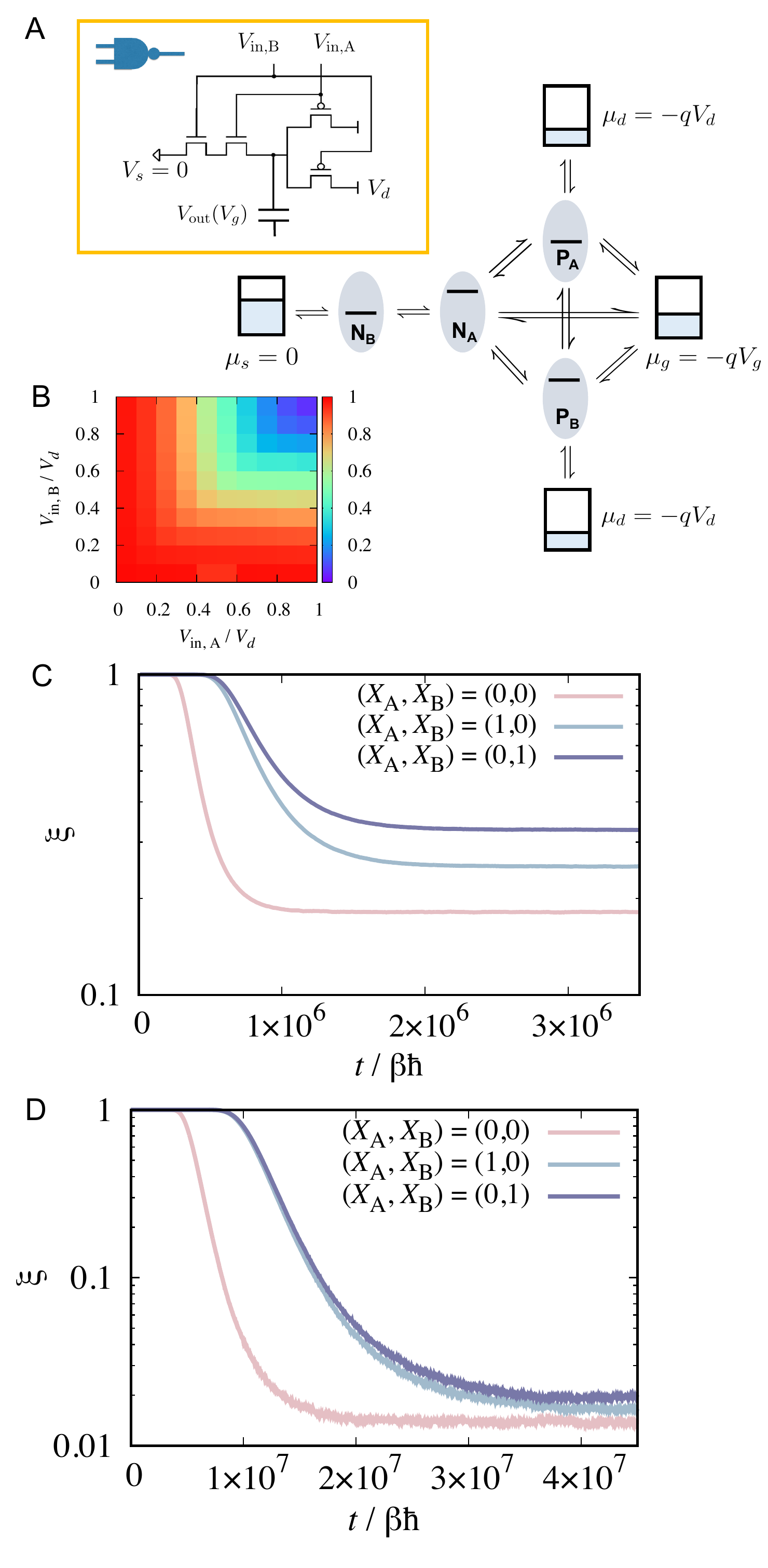}
\caption{Dynamics of a NAND gate. (\textit{A}) Circuit diagram (left) and kinetic diagram (right) of a NAND gate. (\textit{B}) Two-dimensional voltage transfer curve at $V_d=5V_T$. (\textit{C}-\textit{D}) Decay of the error rate with time under three cases:  $(X_{\textrm{in,A}},X_{\textrm{in,B}})=(0,0), (1,0) $ and $(0,1)$ for $V_d=5V_T$ (\textit{C}) and $V_d=8V_T$ (\textit{D}) for a NAND gate initialized with $V_g=0V_T$.}
\label{fig:NAND}
\end{figure}

\subsection{NAND gate}

We have presented a Markovian model for the NOT gate, which reproduces the performance of a perfect gate in the limit of high $V_d$ and for which there is a complex interplay between energy consumption and time. Within the framework presented, it is straightforward to construct an analogous model of a NAND gate. A NAND gate takes in two binary inputs $X_{{A}}$, $X_{{B}}$, and outputs $Y=0$ only when both inputs are 1. As shown in Fig. \ref{fig:NAND}$A$, the kinetic diagram,  similar to the circuit diagram, is composed of two P-type transistors $\textrm{P}_A$, $\textrm{P}_B$, and two N-type transistors $\textrm{N}_A$ and $\textrm{N}_B$. The energy levels of $\textrm{P}_A$ and $\textrm{N}_A$ depend on the first input voltage $V_{\textrm{in},A}$, while the energy levels of $\textrm{P}_B$ and $\textrm{N}_B$ are controlled by the second input $V_{\textrm{in},B}$ (Eq. S14). More details on the model, including the definition of the heat dissipation, can be found in the \textit{SI} (Section IV). The two-dimensional VTC for $V_d=5V_T$ is shown in Fig. \ref{fig:NAND}$B$, which agrees with the truth table for a perfect NAND gate.

While the dynamical properties of the NAND gate are very similar to the NOT gate, an asymmetry arises in the NAND gate due to the different pathways in the kinetic diagram, which is a feature absent in the NOT gate. Consider the three different inputs $(X_{{A}},X_{{B}})=(0,0), (1,0) $ and $(0,1)$ shown in Fig. \ref{fig:NAND}$C$ for $V_d=5V_T$. While for a perfect NAND gate, these three inputs should all correspond to the output $Y=1$, the evolution of the error rate $\xi$ and its converged values in the steady state are not exactly the same for finite $V_d$. In the case of $(X_{{A}},X_{{B}})=(0,0)$, as both $\textrm{P}_1$ and $\textrm{P}_2$ have relatively low energy levels, there are two pathways to charge the capacitor, resulting in a faster error decay rate. For the cases $(X_{{A}},X_{{B}})=(0,1)$ and $(1,0)$, one of the pathways is blocked due to the high energy level of the P transistor, so the error rate decays much slower reflecting the slower charging of the capacitor. While the latter two cases also differ slightly due to the asymmetry in $\textrm{N}_1$ and $\textrm{N}_2$, such differences shrink drastically when we increase $V_d$ to $8V_T$ in Fig. \ref{fig:NAND}$D$. The three cases now converge to similar error rates in the steady state. In fact, as we further increase $V_d$, all such asymmetries vanish, another example of which can be found in Fig. S4, where we plot the one dimensional cut of the VTC along the line $V_{\textrm{in},A}=V_{\textrm{in},B}$ for different $V_d$. As in the case of the NOT gate, our model behaves as a perfect NAND gate as $V_d$ approaches $1e$V. For clarification, we define the propagation delay $\tau_p$ of a NAND gate as the time required to reach the threshold $\alpha V_d$ for the input $(X_A,X_B)=(0,1)$, which is close to $\tau_p$ for the NOT gate of the same $V_d$.

\section{Logic Circuits}

Equipped with a model for the NOT and NAND gates, we now in principle have the tools to implement arbitrary logic functions. While any logic function can be represented in multiple ways, the topology of the circuit has an influence on its accuracy, and thermodynamic costs \cite{wolpert2020thermodynamics}. In the following section, we first explore spatial propagation effects arising from assembling multiple gates in a combinational circuit, and then demonstrate memory effects arising from the feedback loop in a sequential circuit. Understanding the behavior of these basic computing circuits will be crucial to building up a computing device. 

For each logic circuit, which is itself a computing module made up of multiple logic gates, while each gate has an intermediate output, we reserve the symbol $V_{\textrm{out}}$ for the specific $V_g$ that corresponds to the overall output $Y$ of the module. Intermediate input and output voltages are not converted to binaries except for the final output $V_{\textrm{out}}$. 
While the output of each gate is used as the input of the ensuing gate, we neglect the back reaction on $V_{\textrm{out}}$ so that the occupation of the ensuing transistors does not affect $V_{\textrm{out}}$, which is consistent with the high capacitance assumption made in Eq. \ref{eq:epsilon}.
Unless specified otherwise, all gates are initialized at $V_g=0V_T$ at the start of the computation, but no re-initialization is done afterwards. While the channel capacity is a more comprehensive characterization of the accuracy and provides the best case scenario, the much larger input space and complicated memory effects make it cumbersome to calculate in the case of logic circuits. We thus use the error rate in the final output instead, and consider the worst case scenario in choosing the inputs to provide an upper bound for the error rate whenever possible.

\subsection{Combinational Circuit}
\begin{figure}[t]
\centering
\includegraphics[width=0.9\linewidth]{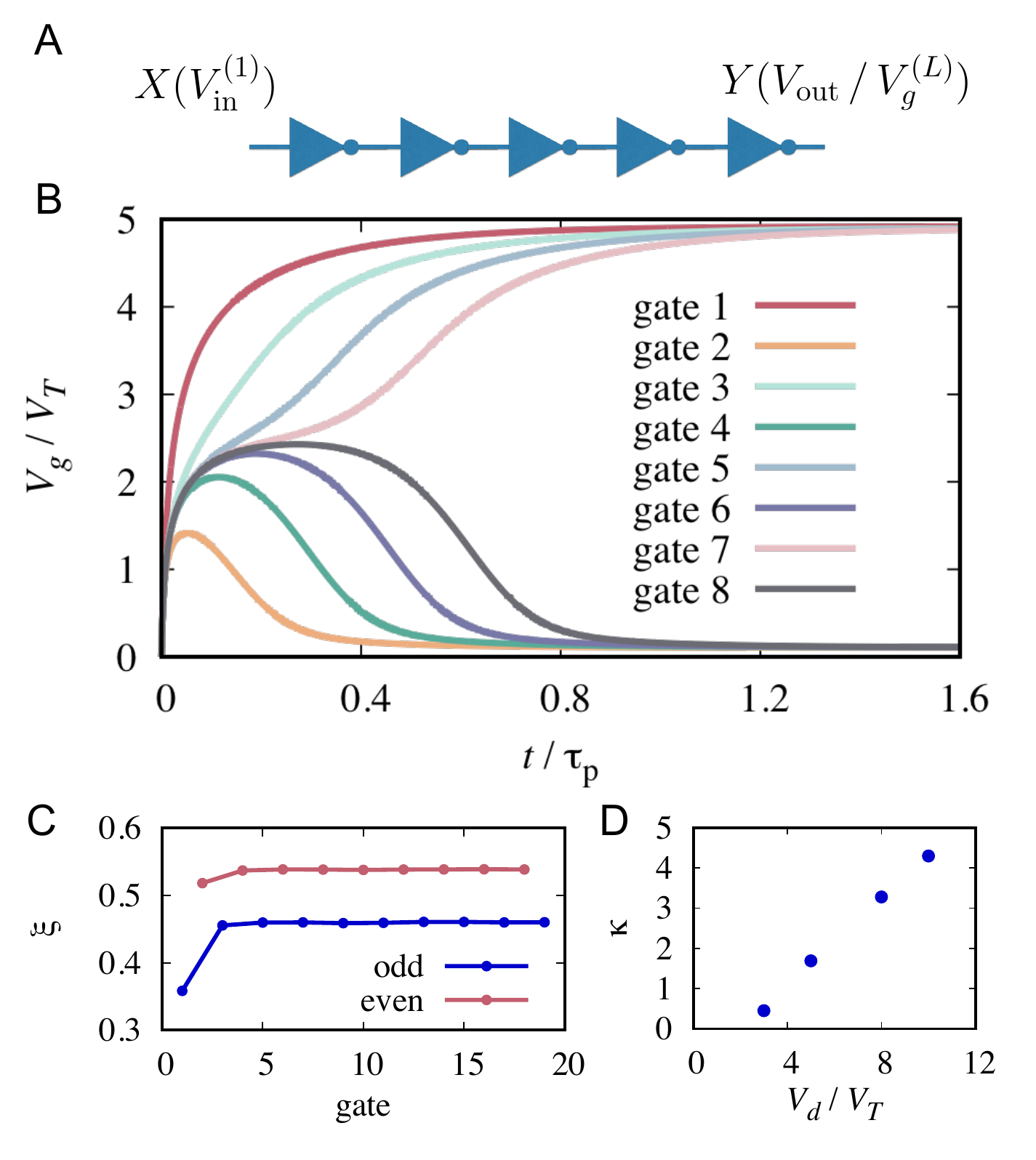}
\caption{Performance of an array of NOT gates with $X=0$, and all $V_g$ initialized to $0V_T$. (\textit{A}) Schematic of an array of NOT gates with a single input $X$ and output $Y$. (\textit{B-C}) Evolution of $V_g$ (\textit{B}), and the steady state error rate (\textit{C}) for individual gates with $V_d=5V_T$. (\textit{D}) Spatial propagation rate $\kappa$ as a function of $V_d$. }
\label{fig:NOTArray}
\end{figure}

A combinational circuit maps a given set of inputs to a single output using a number of gates, such as an adder that computes the sum of inputs and a XOR gate that computes their parity. As the simplest example, we study the behavior of an array of $L$ NOT gates indexed by $i=1,2,\cdots,L$ connected in the way that $V_{\textrm{in}}^{(i)}$ = $V_g^{(i-1)}$ for $i>1$. A schematic of the system can be found in Fig. \ref{fig:NOTArray}$A$. The input of the circuit $X$ determines $V_{\textrm{in}}^{(1)}$, and the output is measured from the last gate $V_{\textrm{out}}=V_g^{(L)}$. The spatial dimension adds complexity to the evolution of $V_g$, as illustrated in Fig. \ref{fig:NOTArray}$B$ for $V_d=5V_T, X=0$. In the steady state, we expect the output voltage of the odd gates close to $V_d$, and the even gates close to zero. For a gate to reach its steady state, its input, which depends on the dynamics of the previous gate, must first reach its expected value, thus the propagation delay should increase with the gate index $i$. As the odd gates are initialized far from their steady state, it will take a significant amount of time to reach its expected output. For the odd gates which have not yet reached the steady state, the ensuing even gate will have a lower input voltage, resulting in the overshoot of voltage before eventually decaying to its expected lower output. The turn over in voltage of the even gates corresponds to the inflection point on the VTC.

A consequence of the connectivity between gates is the corruption of initial input. While the input voltage of the first gate is always $0V_T$, for finite $V_d$, the maximum input voltage of the second gate will be slightly lower than $V_d$, and thus corrupted. As the VTC of the NOT gate is a non-increasing function, a corrupted input will inevitably cause a higher error rate in the output, which will propagate along the array. This is shown in Fig. \ref{fig:NOTArray}$C$, where the error rate for individual gates in the steady state rises initially with gate index, before converging to a constant value after a few gates, and is always higher than that of the single gate. A similar behavior can be found in the propagation delay time, which increases sharply for the first few gates and converges to a slower linear increase afterwards (Fig. S5). This implies that circuit designs with deeper layered structure are unfavorable in terms of both accuracy and propagation delay. 

The convergence behavior is intriguing as it implies the existence of a pair of fixed points $(V^*_{\textrm{odd}},V^*_{\textrm{even}})$ for the intermediate outputs  in the steady state. Indeed, the fixed point solution corresponds to the point on the VTC $(V_{\textrm{in}}=V^*_{\textrm{odd}},V_{\textrm{out}}=V^*_{\textrm{even}})$ satisfying the condition that its reflection $(V_{\textrm{in}}=V^*_{\textrm{even}},V_{\textrm{out}}=V^*_{\textrm{odd}})$ is also on the VTC. As the fixed point is a dynamically stable solution, it does not depend on the initial input $V_{\textrm{in}}^{(1)}$ (Fig. S5), whereas the speed of approaching the fixed point characterizes the spatial correlation in the system. We fit the decay in $|V_g^{(i)}-V^*|/V_T$ with an exponential function $\exp[-\kappa i]$, and report the rate $\kappa$ for different $V_d$ in Fig. \ref{fig:NOTArray}$D$. 
For $V_d=5V_T$, the spatial correlation length $1/\kappa$ is on the order of 1, which means spatial correlation exists between neighboring gates. As a consequence, it is more probable to observe consecutive errors along the array, which is shown by an error analysis of simulated trajectories in the \textit{SI} (Section V). As the VTC becomes sharper with increasing $V_d$,
 the correlation length between gates decreases. In the limit of high $V_d$, the fixed-point solution can be found exactly at $(V_{\textrm{in}}=0,V_{\textrm{out}}=V_d)$, which means that the input becomes uncorrupted. To summarize, the combination of gates introduces longer propagation delay and input corruption, and thus deeper layered circuit design is advised against. By operating at a higher $V_d$ to reduce spatial correlation, the latter problem can be mitigated, but of course this is done at the cost of even longer propagation delay.

\subsection{Sequential Circuit: RS latch}

\begin{figure*}[t]
\centering
\includegraphics[width=17.8cm,height=8.5cm]{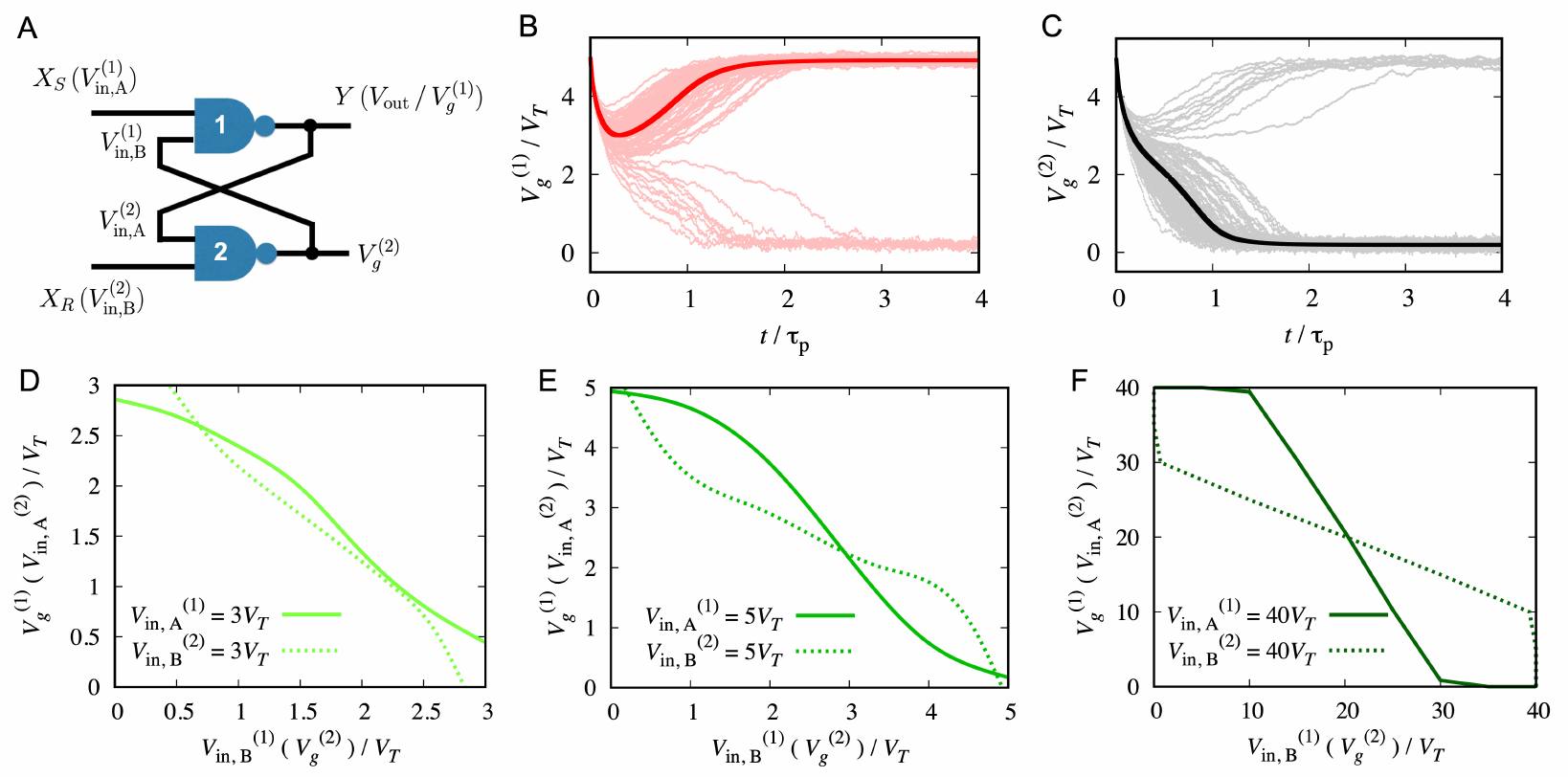}
\caption{ (\textit{A}) Circuit diagram of the RS latch. (\textit{B}-\textit{C}) The evolution of the outputs $V_g^{(1)}$  and $V_g^{(2)}$ for 100 trajectories with the initialization $V_g^{(1)}=V_g^{(2)}=V_d=5V_T$, where time is scaled by the propagation delay $\tau_p$ of the NAND gate. The dark curve represents the average relaxation behavior. (\textit{D}-\textit{F}) The location of the stable informational states determined by overlapping the VTC for the two NAND gates at $V_d=3V_T$ (\textit{D}), $V_d=5V_T$ (\textit{E}), and $V_d=40V_T$ (\textit{F}).}
\label{fig:RSlatch}
\end{figure*}

While combinational circuits are typically used to carry out arithmetic computations, modern computing devices often include another type of logic circuit to handle memory - the sequential circuit. Fig. \ref{fig:RSlatch}$A$ shows an example of such a circuit, known as the RS latch. The RS latch consists of two NAND gates where the output of gate 1, $V_g^{(1)}$, is sent as an input of gate 2, $V_{\textrm{in,A}}^{(2)}$, and similarly, the output of gate 2, $V_g^{(2)}$, is fed back as $V_{\textrm{in,B}}^{(1)}$. 
The remaining two inputs, $V_{\textrm{in,A}}^{(1)}$ and $V_{\textrm{in,B}}^{(2)}$, corresponds to the two external binary inputs $X_\textrm{S}$ and $X_\textrm{R}$, respectively. The output of the circuit, $V_{\textrm{out}}$, which coincides with  $V_g^{(1)}$, depends not only on the external inputs $X_\textrm{S}$ and $X_\textrm{R}$, but also the stored information of $V_g^{(1)}$ and $V_g^{(2)}$. This is the defining characteristic of a sequential circuit, which makes it useful as a memory storage. More specifically, for a perfect RS latch, in the $``$set$"$ stage where the external inputs are set as $X_S=0$, $X_R=1$ or $X_S=1$, $X_R=0$, there exists only one dynamically stable state for the system, so that we can unambiguously designate the memory at $V_{\textrm{out}}$ as 1 or 0. In the $``$hold$"$ stage where $X_S=X_R=1$, however, the system is bistable and its state depends on the initialized value of $V_g^{(1)}$ and $V_g^{(2)}$. In the vicinity of the fixed points, an effective Hamiltonian description of the RS latch is quartic in $V_{\textrm{out}}$ with two minima and a maxima between them \cite{rahman2015rs}. This emergent bistability resulting from the feedback loop allows the RS latch to function as a memory storage device. 

To function as a memory storage device, a circuit must have at least two distinguishable states in which information can be stored. For our stochastic model in Fig. \ref{fig:RSlatch}$A$, these states correspond to the steady-state solutions that satisfy the feedback condition $V_{\textrm{in,B}}^{(1)}=V_g^{(2)}$, $V_{\textrm{in,A}}^{(2)}=V_g^{(1)}$ under the input  $V_{\textrm{in,A}}^{(1)}=V_{\textrm{in,B}}^{(2)}=V_d$. An intuitive way to find their location is to overlap the VTC of the two NAND gates along the cut $V_{\textrm{in,A}}^{(1)}=V_d$ and $V_{\textrm{in,B}}^{(2)}=V_d$, which are not exactly the same due to the asymmetry in the non-perfect NAND gates. We show a couple of scenarios at different $V_d$ in Figs. \ref{fig:RSlatch}$D$-$F$. At $V_d=3V_T$, the highly asymmetric VTCs cross merely at $(V_{\textrm{in,B}}^{(1)},V_{\textrm{in,A}}^{(2)})=(0.67V_T, 2.61V_T)$, indicating that the system only has a single stable state and does not qualify as a memory storage device. As $V_d$ increases to $5V_T$, two dynamically stable informational states start to emerge at $(V_{\textrm{in,B}}^{(1)},V_{\textrm{in,A}}^{(2)})=(0.19V_T, 4.92V_T)$ and $(4.89V_T, 0.20V_T)$, though the slight asymmetry suggests different dynamics around the two states. While a third intersection point is found at $(V_{\textrm{in,B}}^{(1)},V_{\textrm{in,A}}^{(2)})=(2.93V_T, 2.26V_T)$, it corresponds to an unstable saddle point. At an even higher $V_d=40V_T$, the two states converge to $(V_{\textrm{in,B}}^{(1)},V_{\textrm{in,A}}^{(2)})=(0V_T, 40V_T)$ and $(40V_T, 0V_T)$, and symmetry is restored.

While the existence of two distinguishable informational states is guaranteed at sufficiently high $V_d$, there remains the question of whether these informational states are robust against noises. While in both the set and hold stages, $V_g^{(1)}$ and $V_g^{(2)}$ are usually sufficiently far from each other that it is possible to distinguish them definitively, there do exist occasions where the noise can mediate a transition. One such example is shown in Figs. \ref{fig:RSlatch}$B$ and $C$ for the initialization $V_g^{(1)}=V_g^{(2)}=V_d=5V_T$. As the outputs of the gates evolve from their initialization towards the steady state solution, there is a significant overlap between the two outputs around $t=0.5\tau_p$, which leads to about 13$\%$ percent of the trajectories failing to retain the information and evolving to the wrong fixed point. This kind of perturbation happens when the overlap region includes the unstable intersection point on the VTC, and the change of convexity of the effective Hamiltonian brings the trajectory towards a different stable state. Such an initialization error is rare to observe either in the set or hold stage, and we show additional evidence for the robustness of the circuit at $V_g^{(1)}=V_g^{(2)}=2.5V_T$ and $0V_T$ in Fig. S7. In addition, at a higher $V_d$, as the VTC becomes sharper, not only do the two minima in the Hamiltonian become more separated, their vicinity also become steeper, both of which facilitate the differentiation between the two states and thus will drastically improve the robustness of the device.

\subsection{Sequential Circuit: D flip-flop}

\begin{figure}[t]
\centering
\includegraphics[width=0.92\linewidth]{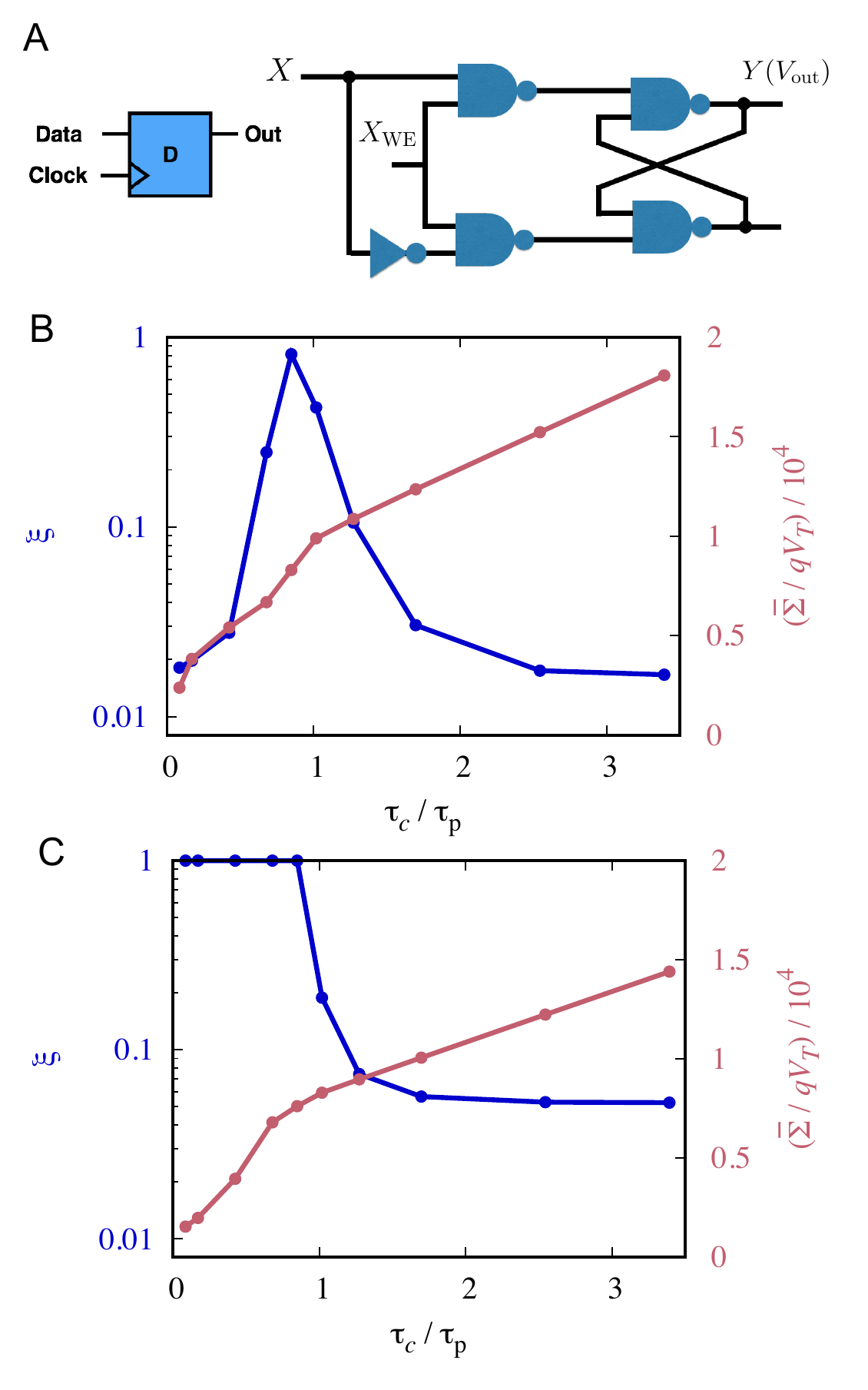}
\caption{ (\textit{A}) The symbol (left) and circuit diagram (right) of a D flip-flop. (\textit{B}-\textit{C}) Error rate (blue, with axis label on the left) and average dissipation per gate per cycle (red, with axis label on the right) as a function of the clock cycle $\tau_c$, scaled by the propagation delay of the NAND gate, for cycles with input $X=1$ (\textit{B}) and $0$ (\textit{C}). All NAND gates are operated at $V_d=8V_T$.}
\label{fig:Dflipflop}
\end{figure}

With the RS latch as a basic computing unit, we can model a memory storage module that synchronizes with the clock generator, called the D flip-flop. Modern computing devices typically include a pulse generator that oscillates between 0 and 1, with a clock cycle $\tau_c$. To see how the clock is incorporated into the D flip-flop, we show the circuit diagram of a D flip-flop in Fig. \ref{fig:Dflipflop}$A$, built up from 4 NAND gates and 1 NOT gate. 
The circuit can be readily modularized as a memory storage unit, denoted with the symbol D,  that takes in an input  $X$ representing the data, another input $X_{\textrm{WE}}$ synchronized with the clock, and generates an output $Y$. The two NAND gates with the feedback loop on the right hand side constitute an RS latch, which is responsible for the memory storage. When the write-enable input  $X_{\textrm{WE}}=1$, the D flip-flop sets its output $V_{\textrm{out}}$ in agreement with the data $X$, whereas when $X_{\textrm{WE}}=0$, the D flip-flop holds its stored value as its output, which can be further processed for computing purposes. 

The clock cycle $\tau_c$, or the clock frequency $1/\tau_c$, is an important parameter as it determines how fast data can be read and stored. In Figs. \ref{fig:Dflipflop}$B$ and $C$ we illustrate how the clock cycle influences the accuracy and dissipation of the data transmission process for a D flip-flop with $V_d=8V_T$. We start with $X_{\textrm{WE}}=1$ and send in a stream of data $\textbf{X}=\{1,0,1,0,\cdots\}$. While $X_{\textrm{WE}}$ alternates between 1 and 0 every $\tau_c/2$, the data input only changes every $\tau_c$. This input data sequence is chosen to maximize the alternation in the output, and thus minimize the memory effect discussed earlier for the NOT gate. Therefore, the error rate and dissipation in this case are expected to be the highest among all possible input sequences.  The error rate $\xi$ is measured according to the output $V_{\textrm{out}}$ at the end of each cycle,  and is reported separately for the cycles with $X=1$ and 0. 
The evolution of $V_{\textrm{out}}$ as a function of the cycle number can be found in Fig. S8.
 
Similar to the behavior for the single NAND gate in Fig. \ref{fig:NAND}$D$, the error rate for $X=0$ starts to decrease monotonically when $\tau_c$ is longer than the single gate propagation delay $\tau_p$. The error rate for $X=1$, however, first increases with $\tau_c$ before eventually decreasing. This counter-intuitive behavior comes from the memory retention behavior in the RS latch. Once data are stored in the RS latch, 
it tends to stay in the memory by influencing the transmission of the following data, and thus introduces temporal correlation between the outputs. The influence of the data can only be erased given sufficient time to transmit the following data. This temporal correlation time, or memory retention time, again coincides with the propagation delay $\tau_p$. In this example, as the first input $X=1$, the output retains the memory of a higher output at short $\tau_c$, so the error rate for $X=1$ is deceptively low, and the error rate for $X=0$ is high. At $\tau_c\approx\tau_p$, the output is stuck between the high and low outputs before reaching either steady state, so that the error rate for either cycle is high. In this regime, the average dissipation accumulated within each cycle rises fast with $\tau_c$, as charging processes contribute heavily to energy costs. When $\tau_c>\tau_p$, the memory effect is eventually overcome and the error rates for both cycles start to decay. The average dissipation rate also converges to a smaller constant value as within each cycle, the system is able to reach the steady state, in which much less dissipation is generated. The exponential scaling of $\tau_p$ with $V_d$ implies that while the asymptotic error is expected to decrease when operating far from thermal energies $V_d\gg V_T$, the speed with which the D flip-flop can function with that lower error is significantly slower. Due to this lag, comparing between a lower and a higher $V_d$, the error is expected to be much lower in the former case for a fixed computing time on the order of $\tau_p$ of the lower $V_d$. 

\section{Parity Computing Device}

With the combinational circuit modularized as the arithmetic logic unit (ALU), and the sequential circuit as the memory storage device, we can combine the two components to model a computing device. We choose the task of computing the parity of a sequence of inputs $\textbf{X}=\{X_1,X_2,\cdots,X_{N}\}$ of length $N$, which has wide applications in error detection. Such a task can be easily implemented by combining $(N-1)$ XOR gates in a sequential manner. However, when $N$ is relatively large, due to the limitation in resources, it is beneficial to break up the task in several steps, and store intermediate results in  memory. The clock generator synchronizes the operation of different components to ensure correct sequencing.

\begin{figure}[t]
\centering
\includegraphics[width=0.95\linewidth]{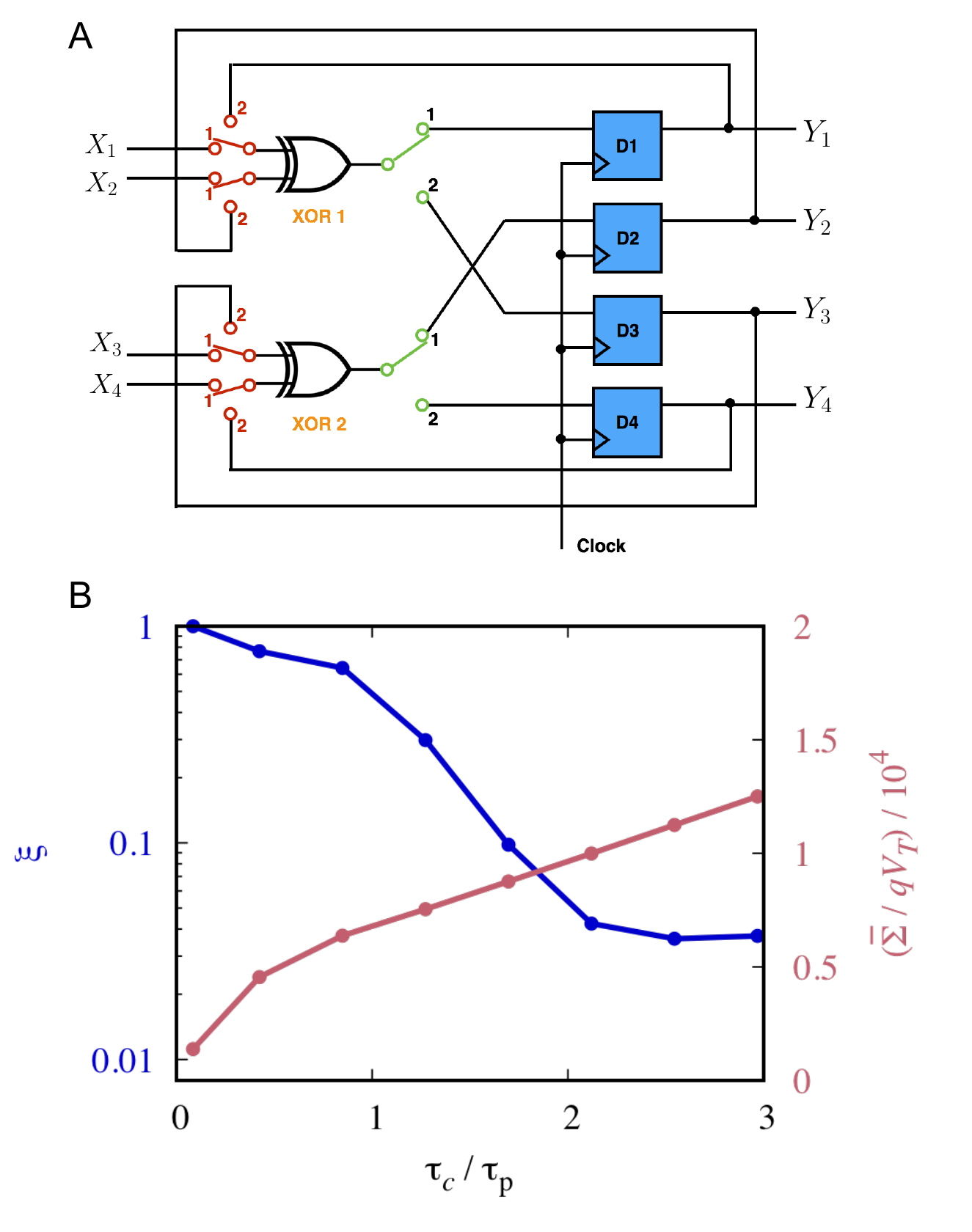}
\caption{ (\textit{A}) Schematic of the parity computing device with 2 XOR gates and 4 D flip-flops. The input of each XOR gate is controlled by 2 input two-way switches, shown in red. The output two-way switch, shown in green, determines which D flip-flop is used to store the output of the XOR gate. (\textit{B}) The average error rate (blue, with axis label on the left) and dissipation per gate per cycle (red, with axis label on the right) as a function of the time cycle $\tau_c$, scaled by the propagation delay of a single NAND gate, averaged over different input sequences. All gates are operated at $V_d=8V_T$. }
\label{fig:device}
\end{figure}

As an example,  we consider two XOR gates as an ALU, and four D flip-flops, D1 to D4, as a memory device to check the parity of $N=12$. Fig. \ref{fig:device}$A$ shows the schematic of our design, while the complete circuit diagram can be found in Fig. S9.  Each XOR gate takes in 2 binary inputs at a time, the source of which is controlled by 2 input two-way switches, shown in red in Fig. \ref{fig:device}$A$. When the switch is connected to terminal 1, the input comes from the data sequence $\textbf{X}$; whereas when terminal 2 is connected, the input comes from the data stored in a D flip-flop. At the end of each XOR gate is an output two-way switch, shown in green in Fig. \ref{fig:device}$A$, which controls where to store the output. We store new data only on free D flip-flops, where the data stored at an earlier time is already read out for post-processing and does not need to be held any more. The total system requires modeling over 100 transistors.

We start the computation by sending in pairs of input data from the data sequence, and computing their parities with the XOR gates. The D flip-flops are set by outputs from the ALU (first D1, D2 and then D3, D4), and once all D flip-flops have been set, we free them by sending the stored information back to the ALU for further processing. The computation is terminated when all inputs are taken into account in the final output, and the entire task can be completed in 6 clock cycles. A more detailed description of the protocol, and a computational tree graph that illustrates how intermediate outputs are related to the final output can be found in the $\textit{SI}$ (Section VI) and Fig. S10.

As before, we are interested in the time and dissipation required to achieve a certain accuracy. In Fig. \ref{fig:device}$B$, we show the error rate for the final output at $t=6\tau_c$, and the average dissipation per gate (averaged over the 28 gates in this device) per clock cycle $\bar{\Sigma}$ as a function of $\tau_c$ with $V_d=8V_T$. Both results are averaged over more than $10^4$ inputs, which are sequences of independent and identically distributed Bernoulli random variables with equal probability of being 0 or 1. As expected, the average error rate decays with the clock cycle until $\tau_c\approx 3\tau_p$, as the extended spatial dimension of the circuit increases the propagation delay in the final output. At such a high $V_d$, spatial correlations do not extend beyond neighboring gates, and are even weaker between different modules, especially for clock cycles longer than $\tau_p$. We further analyze how the error in the final output is correlated along its computational path in the \textit{SI}. The average dissipation first increases sharply and then converges to a linear growth in the limit of large $\tau_c$, similar to the D flip-flop, but slightly lower than that in Fig. \ref{fig:Dflipflop}$C$ for the same $\tau_c$. This is because the input sequences are randomly chosen instead of alternating between 0 and 1, and the memory effect can help shorten the charging process, which most contributes to the entropy production. Additionally, because of the synchronization, the D flip-flops may remain at a steady state for a few cycles before they are freed to store new data. During such periods, the dissipation is especially low as the entropy production in the steady state is minimal due to relatively small leakage currents. Therefore, computational protocols that minimize changes on the memory storage device are desirable for low dissipation computing. Taking into consideration both the accuracy and dissipation, the optimal clock cycle to operate with is $\tau_c\approx2\tau_p$, as lowering the speed further will only result in higher dissipation from the steady state.

\section{Discussion and Conclusion}

We have illustrated a promising model for stochastic logic gates, and demonstrated its utility in building arbitrary logical circuits. Information manipulations, such as bit storage and erasing, are represented by the charging and discharging of the capacitors, which is consistent with current data storage technology. 
While our model performs as a perfect logic circuit when operated in the limit of low noise, its thermodynamical consistency allows us to study the rich interplay between speed, accuracy, and dissipation in the intermediate regimes, from which we can derive some useful design principles for low dissipation computing devices. For instance, we have provided a physical origin of input corruption in the combinational circuits, as well as feedback robustness in the sequential circuits, and illustrated how each can be improved drastically by operating at a slightly higher voltage. In addition, memory effects should be exploited as much as possible to minimize dissipation. 
With modularization, it is straightforward to scale up our model to even larger and more complex systems, making it a useful model to study collective behaviors of circuits.  It is useful to bear in mind that the signal-to-noise ratio regime that is explored in this work is two orders of magnitude lower than current technology. However, with the exponential growth of the number of computations per unit of energy dissipated, as observed by Koomey's law \cite{koomey2010implications}, such a low dissipation regime will soon become relevant. While the model we propose is not intended for a direct comparison with the current CMOS technology, the fact that it can reproduce the input-output behaviors of universal logic gates makes it a promising tool to study fundamental physical limits on computations.

One of the major motivations of this work is to enable the design of low dissipation computing devices with maximal accuracy and speed. While there exist several theoretical results that propose bounds on the thermodynamic costs of computing \cite{wolpert2019stochastic,wolpert2020thermodynamics}, understanding under what circumstances they are saturated requires a realistic model for the thermal noise. 
As each dynamical process in our model obeys a local detailed balance, we are able to harness the lessons of stochastic thermodynamics to define and analyze the time dependence and fluctuations of the entropy production. Note that the $\Sigma$ we have referred to throughout the paper is different from the total dissipation, which is the heat released by the system, by a term $T\Delta S$, the change in the Shannon entropy of the system transistors times the bath temperature. Nevertheless, we have used the two terms interchangeably since for the timescales studied, the boundary term $\Delta S$ is orders of magnitude smaller than the cumulative term $\Sigma$, which is very large due to the large gate capacitance. This then raises the question of how to further decrease the irreversible dissipation and that associated with charging the gates. This problem is the crux of optimal control theory, and adiabatic circuit design~\cite{frank2003common,zulehner2019design}, from which some design principles can be borrowed. For example, while we have kept the input voltage of the transistors $V_{\textrm{in}}$ fixed within each cycle, one can design optimal feedback protocol that controls it according to the state of the capacitor, in order to minimize the irreversible dissipation throughout the process. Such optimal feedback protocols already exist for simple thermodynamic engines~\cite{horowitz2011designing}, and we believe our model provides an ideal testing ground for applying more advanced stochastic control algorithms~\cite{das2021variational}. Marrying our model with a framework that integrates information with thermodynamics \cite{deffner2013information,barato2014stochastic}, we hope to get a step closer to achieving a computing design that minimizes dissipation while maximizing accuracy and speed. 

\section{Materials and Methods}
Simulations were done with both an iterative, numerically exact diagonalization of the master equation as well as Gillespie simulations \cite{gillespie1976general}. In both, we employ a separation of timescales for electron transfer to or from a transistor and gate charging, afforded by the large gate capacitance. Specifically, the large capacitance means we can update $V_g$ with discrete time-step, chosen to be 10$\beta \hbar$, and compute rates at fixed $V_g$ in-between these dynamical updates.  More details on the models and calculations can be found in the $\textit{SI}$. All our codes and data can be accessed on GitHub: https://github.com/chloegao12.

\section{Acknowledgement}
The authors thank Gavin E. Crooks for suggesting this topic for study and for invaluable discussion and comments. C.Y.G. and D.T.L. are  supported by the US Department of Energy, Office of Basic Energy Sciences, through the CPIMS Program Early Career Research Program under Award DE-FOA0002019.

\section{References}
\bibliography{pnas-sample}

\end{document}



\title{Supplementary Information for Principles of Low Dissipation Computing from a Stochastic Circuit Model}
\author{Chloe Ya Gao}
\affiliation{Department of Chemistry, University of California, Berkeley, California \looseness=-1}
\author{David T. Limmer}
\email{dlimmer@berkeley.edu}
\affiliation{Department of Chemistry, University of California, Berkeley, California \looseness=-1}
\affiliation{Kavli Energy NanoScience Institute, Berkeley, California \looseness=-1}
\affiliation{Materials Science Division, Lawrence Berkeley National Laboratory, Berkeley, California \looseness=-1}
\affiliation{Chemical Science Division, Lawrence Berkeley National Laboratory, Berkeley, California \looseness=-1}

\date{\today}

\maketitle

\section{NOT gate}
The Markovian system is described by the occupation number of the two single-electron levels $n_{\textrm{N}}$, $n_{\textrm{P}} =0/1$, and the gate voltage $V_g$. Electrons can jump between the transistors and the reservoirs only if the target site is empty. Denoting the transition rate from state $i$ to $j$ as $k_{ji}$, the rates describing the exchange of electrons between the transistors and the reservoirs are given by
\begin{equation}
\begin{aligned}
k_{\textrm{N}s}&=\Gamma f_s(\epsilon_\textrm{N}),\hspace{5mm}
k_{s\textrm{N}}=\Gamma[1-f_s(\epsilon_{\textrm{N}})],\\
k_{\textrm{P}d}&=\Gamma f_d(\epsilon_{\textrm{P}}),\hspace{5mm} 
k_{d\textrm{P}}=\Gamma[1-f_d(\epsilon_{\textrm{P}})],\\
k_{\textrm{N}g}&=\Gamma f_g(\epsilon_{\textrm{N}}),\hspace{5mm} 
k_{g\textrm{N}}=\Gamma[1-f_g(\epsilon_{\textrm{N}})],\\
k_{\textrm{P}g}&=\Gamma f_g(\epsilon_{\textrm{P}}),\hspace{5mm} 
k_{g\textrm{P}}=\Gamma[1-f_g(\epsilon_{\textrm{P}})],\\
\end{aligned}
\label{eq:k_tr}
\end{equation}
where $f_i(x)=[e^{\beta(x-\mu_i)}+1]^{-1}$ is the Fermi distribution. The transition rate between the two transistors depends on their relative energy levels, for example, in the case of $\epsilon_{\textrm{P}}>\epsilon_{\textrm{N}}$,
\begin{equation}
k_{\textrm{PN}}=\Gamma n(\epsilon_{\textrm{P}}-\epsilon_{\textrm{N}}),\hspace{5mm} k_{\textrm{NP}}=\Gamma[1+n(\epsilon_{\textrm{P}}-\epsilon_{\textrm{N}})],
\label{eq:k_tt}
\end{equation}
where $n(x)=[e^{\beta x}-1]^{-1}$ is the Bose-Einstein distribution. The rate constant $\Gamma=0.2/\beta\hbar\sim \textrm{ps}^{-1}$ is chosen so that electron transitions happen on a longer timescale than quantum tunneling, and the broadening of energy levels due to the coupling to electrodes is smaller than thermal fluctuations.

The dynamics of the capacitor is solved by the equation of motion
\begin{equation}
dV_g=-\frac{1}{C_g}\int_0^{t_{\textrm{int}}}J_g(t)dt,
\label{eq:Cg}
\end{equation}
where $C_g$ is the capacitance and $J_g$ is the electron current flowing into the electrode from the transistors. While the transfer of electrons changes $V_g$, the capacitor is treated as an electron reservoir at constant chemical potential $\mu_g=-qV_g$ within each time interval $t_{\textrm{int}}$. Thus, the assumption made here is that the electron transfer within each $t_{\textrm{int}}$ is small compared to $C_gV_T$, and the electron relaxation within the capacitor is fast compared to $t_{\textrm{int}}$. We have chosen $C_g=200q/V_T$, $t_{\textrm{int}}=10\beta\hbar$ in order to justify these assumptions. 

To obtain a numerically exact solution to the Markovian dynamics, for each interval $t_{\textrm{int}}$, we solve for the average occupation number $\left\langle n_{\textrm{N}} \right\rangle$, $\left\langle n_{\textrm{P}} \right\rangle$ from the stationary solution of the master equation, which describes how the probability of the configuration $p_{n_{\textrm{N}},n_{\textrm{P}}}$ evolves with time:
\begin{equation}
\begin{bmatrix}
\dot{p}_{0,0}\\ \dot{p}_{0,1}\\ 
\dot{p}_{1,0}\\ \dot{p}_{1,1}
\end{bmatrix}=\begin{bmatrix}
-S_1 & k_{r\textrm{P}}+k_{g\textrm{P}} & k_{l\textrm{N}}+k_{g\textrm{N}} & 0 \\
k_{\textrm{P}r}+k_{\textrm{P}g} & -S_2 & k_{\textrm{P}\textrm{N}} & k_{l\textrm{N}}+k_{g\textrm{N}}\\ 
k_{\textrm{N}l}+k_{\textrm{N}g} & k_{\textrm{N}\textrm{P}} & -S_3 & k_{r\textrm{P}}+k_{g\textrm{P}}\\ 
0 & k_{\textrm{N}l}+k_{\textrm{N}g} & k_{\textrm{P}r}+k_{\textrm{P}g} & -S_4
\end{bmatrix}
\begin{bmatrix}
p_{0,0}\\ p_{0,1}\\ p_{1,0}\\ p_{1,1}
\end{bmatrix}
\label{eqn:NOT_master}
\end{equation} 
where $S_j=\sum_{i\neq j}W_{ij}$ for a matrix $W$. Note that the transition rates concerning the gate ($g$) are time dependent through $V_g$, while a local equilibrium approximation is invoked within each integration interval  $t_{\textrm{int}}$. The current $J_g$ flowing into the capacitor is then computed by the sum of two terms
\begin{equation}
\begin{aligned}
J_{\textrm{N}\rightarrow g}/q&=k_{g\textrm{N}}\left\langle n_{\textrm{N}} \right\rangle-k_{\textrm{N}g}(1-\left\langle n_{\textrm{N}} \right\rangle),\\
J_{\textrm{P}\rightarrow g}/q&=k_{g\textrm{P}}\left\langle n_{\textrm{P}} \right\rangle-k_{\textrm{P}g}(1-\left\langle n_{\textrm{P}} \right\rangle).
\end{aligned}
\label{eq:Jg}
\end{equation}

In the Gillespie simulation, the electron jumping processes are modeled explicitly as chemical reactions, with $M=10$ reaction rates
\begin{equation}
\begin{aligned}
w_1&=k_{\textrm{N}s}(1-n_{\textrm{N}}),\hspace{10.4mm}
w_2=k_{s\textrm{N}}n_{\textrm{N}},\\
w_3&=k_{\textrm{P}d}(1-n_{\textrm{P}}),\hspace{10.6mm}
w_4=k_{d\textrm{P}}n_{\textrm{P}},\\
w_5&=k_{\textrm{N}g}(1-n_{\textrm{N}}),\hspace{10.2mm}
w_6=k_{g\textrm{N}}n_{\textrm{N}},\\
w_7&=k_{\textrm{P}g}(1-n_{\textrm{P}}),\hspace{10.6mm}
w_8=k_{g\textrm{P}}n_{\textrm{P}},\\
w_9&=k_{\textrm{PN}}n_{\textrm{N}}(1-n_{\textrm{P}}),\hspace{5mm}w_{10}=k_{\textrm{NP}}n_{\textrm{P}}(1-n_{\textrm{N}}).
\end{aligned}
\end{equation}
We use the Monte Carlo method to simulate the probability that reaction $i$ will happen after time $t$,
\begin{equation}
P(t,i)=w_i\exp[-\sum_{i=1}^{M}w_it],
\end{equation}
and the currents between two sites are calculated as the discrete number of jumps between the two sites. The discretization error in voltages between the average protocol and the Gillespie simulation is on the order of $q/C_g=0.005V_T$. 

\section{Computation of channel capacity}
For the NOT gate, we observe Gaussian distributions in $V_{\textrm{out}}$ in the steady state regardless of $V_d$, with the same variance $1/(\beta C_g)$. Under the Gaussian assumption, the error rate is uniquely determined by the average output voltage $V_{\textrm{out}}$. For example, the accuracy rate for $X=0$ is determined by the conditional probability
\begin{equation}
p(Y=1|X=0)=\int_{0.98V_d}^{\infty}dV\,\,\sqrt{\frac{\beta C_g}{2\pi }}\times\exp\left[-\frac{\beta C_g(V-V_{\textrm{out}})^2}{2}\right].
\end{equation}
Marginalizing the conditional probabilities, the mutual information can be computed by
\begin{equation}
I(X;Y)=\sum_{x=0,1}\sum_{y=0,1,\varnothing}p(x,y)\log_2\frac{p(x,y)}{p(x)p(y)}.
\end{equation}
To compute the channel capacity, we numerically maximize $I(X;Y)$ over the base probability distribution $p(X)$, where the conditional probabilities are computed using $V_{\textrm{out}}$ in the steady state.

\section{Computation of average information rate}
For an information channel with memory, the average information rate per data is defined as
\begin{equation}
\bar{I}(\textbf{X};\textbf{Y})=\frac{1}{N_{\textrm{data}}}I(X_1,\cdots,X_{N_{\textrm{data}}};Y_1,\cdots,Y_{N_{\textrm{data}}}),
\end{equation}
which in the limit of $N_{\textrm{data}}\rightarrow\infty$ and upon maximizing over the input probability distribution $p(\textbf{X})$, is the generalization of the channel capacity. As an example, for $N_{\textrm{data}}=2$, the mutual information is computed by
\begin{equation}
\begin{aligned}
I(X_1,X_2;Y_1,Y_2)&=\sum_{x_1=0,1}\sum_{x_2=0,1}\sum_{y_1=0,1,\varnothing}\sum_{y_2=0,1,\varnothing}\\
&p(x_1,x_2,y_1,y_2)\log_2\frac{p(x_1,x_2,y_1,y_2)}{p(x_1,x_2)p(y_1,y_2)}.
\end{aligned}
\end{equation}
To incorporate the memory effect, note that the probability distribution of the $i$th output $Y_i$ is not only a function of $X_i$, but also the output voltage of the previous data $V_{\textrm{out}}^{i-1}$. The dependence can be expressed with the conditional probability $p(V_{\textrm{out}}^i|x_i,V_{\textrm{out}}^{i-1})$, which we sample by Gillespie simulations of more than $10^7$ trajectories. The dependence of $I$ on the observation time $\tau_{\textrm{obs}}$ thus comes from this conditional probability.

Let $V_{\textrm{out}}^0=0V_T$, and we can write the joint probability
\begin{equation}
p(x_1,x_2,V_{\textrm{out}}^1,V_{\textrm{out}}^2)=p(V_{\textrm{out}}^2|x_2,V_{\textrm{out}}^1)p(V_{\textrm{out}}^1|x_1,V_{\textrm{out}}^0=0V_T)p(x_2)p(x_1),
\end{equation}
which follows from the Markovian nature of the memory effect, and the fact that the input data $x_i$ are independent from each other. The joint distributions can then be computed by marginalization, e.g., defining the mapping between $Y$ and $V_{\textrm{out}}$ in Eq. 5 as $Y=\textrm{dig}(V_{\textrm{out}})$,
\begin{equation}
p(x_1,x_2,y_1,y_2)=\sum_{y_1=\textrm{dig}(V_{\textrm{out}}^1)}\sum_{y_2=\textrm{dig}(V_{\textrm{out}}^2)} p(x_1,x_2,V_{\textrm{out}}^1,V_{\textrm{out}}^2),
\end{equation}
where the sum is over all $V_{\textrm{out}}^i$, discrete in our simulation, that correspond to $y_i$.

As the numerical maximization is difficult for large $N_{\text{data}}$, without loss of generality, we choose as our input a sequence of independent and identically distributed Bernoulli random inputs with equal probability of being 0 or 1. We show in Fig. S2 the average information rate at $V_d=5V_T$ for $N_{\textrm{data}}=1,2,3$ as a function of $\tau_{\textrm{obs}}$, the processing time for each individual data from input to output. For $N_{\textrm{data}}=1$, the information rate first decreases at small $\tau_{\textrm{obs}}$, as $\xi(X=1)$ inevitably increases at short time due to the initialization $V_g=0V_T$, and rises up sharply around the propagation delay $\tau_p$, which is the time required for $\xi(X=0)$ to decay. As we increase $N_{\textrm{data}}$, the memory effect is expected to be especially helpful when consecutive inputs share the same value, and thus should on average improve the information rate. This effect is not evident for extremely small $\tau_{\textrm{obs}}$, where the error rate for $X=0$ is too high to be corrected by the memory effect. However, the memory effect plays a significant role, bringing up to 30 percent increase in the average information rate, at intermediate $\tau_{\textrm{obs}}$. All curves eventually converge in the long time limit as the memory effect wears off for times much longer than the propagation delay. 

\section{NAND gate}
The NAND gate is described by four occupation numbers $(n_{\textrm{P}_A}, n_{\textrm{P}_B}, n_{\textrm{N}_A}, n_{\textrm{N}_B})$ and the gate voltage $V_g$. The energy levels of the four transistors are determined in the same manner as in Eq. 2 of the main text, with $\epsilon_{\textrm{N}_A}$, $\epsilon_{\textrm{P}_A}$ corresponding to $V_{\textrm{in,A}}$, and $\epsilon_{\textrm{N}_B}$, $\epsilon_{\textrm{P}_B}$ corresponding to $V_{\textrm{in,B}}$,
\begin{equation}
\begin{aligned}
&\epsilon_{\textrm{P}_A}=\epsilon_{\textrm{P}}^{0} + qV_{\textrm{in},A},\,\quad \epsilon_{\textrm{N}_A}=\epsilon_{\textrm{N}}^{0} - q V_{\textrm{in},A},\\
&\epsilon_{\textrm{P}_B}=\epsilon_{\textrm{P}}^{0} + qV_{\textrm{in},B},\,\quad \epsilon_{\textrm{N}_B}=\epsilon_{\textrm{N}}^{0} - q V_{\textrm{in},B},
\end{aligned}
\end{equation}
The transition rates between the transistors and reservoirs are described analogously to Eqs. ~S\ref{eq:k_tr} and ~S\ref{eq:k_tt}. To avoid numerical issues in Eq. ~S\ref{eq:k_tt} when the adjacent transistors have the same energy levels, we add a $10^{-3}$ regularizer in the denominator of the Bose-Einstein distribution $n(x)$. The output $V_g$ is again treated as a capacitor described by Eq.~S\ref{eq:Cg}, where the current $J_g$ is the sum of $J_{\textrm{P}_A\rightarrow g}$, $J_{\textrm{P}_B\rightarrow g}$, and $J_{\textrm{N}_A\rightarrow g}$, each defined as in Eq.~S\ref{eq:Jg}. As in the NOT gate, both an average protocol and a Gillespie simulation consisting of 16 chemical reactions are used to study the dynamics. The heat dissipation during an observation time $\tau_{\textrm{obs}}$ is 
\begin{equation}
\Sigma(\tau_{\textrm{obs}})=\int_0^{\tau_{\textrm{obs}}} dt\,\, J_{s\rightarrow \textrm{N}_B}(\mu_s-\mu_g)+(J_{d\rightarrow\textrm{P}_A}+J_{d\rightarrow\textrm{P}_B})(\mu_d-\mu_g).
\end{equation}

\section{Error analysis in an array of NOT gates}

We simulate an array of NOT gates of length $L$ with $V_{\textrm{in}}^{(1)}=0V_T$, $V_d=5V_T$, and generate more than $10^8$ snapshots of the system. For each snapshot, we first search for regions with $d_e=1,2,\cdots,16$ consecutive errors, and then count the total number of such error domains, denoted by $N(d_e)$. While counting, we do not account for the first 10 gates in each array as they have not reached the fixed point solution. To characterize spatial correlation in the system, we compare the value of $N(d_e+1)/N(d_e)$ computed in our model with the case where all gates are independent from each other. We denote the single gate error rate of the odd and even gates as $\xi_{0/1}$. Note that to make a fair comparison, this error rate corresponds to the fixed point solution of the array, instead of the error rate of a single NOT gate with $X=0/1$. Assuming odd and even gates are observed with equal probability, it is easy to derive that $N(d_e=1)=(\xi_0+\xi_1)/2$ and $N(d_e=2)=\xi_0\xi_1$. One can infer from this simple calculation that for independent gates, $N(d_e+1)/N(d_e)=2\xi_0\xi_1/(\xi_0+\xi_1)$ if $d_e$ is odd, $(\xi_0+\xi_1)/2$ if $d_e$ is even. This result is plotted in Fig. S6 as the reference, where the zigzag behavior comes from the difference between $\xi_0$ and $\xi_1$. In addition, we plot in the same figure the value $N(d_e+1)/N(d_e)$ for our model with $L=60,110,160$ and 210. For smaller $L$, as finite size effect prevents larger error domains to emerge, the value $N(d_e+1)/N(d_e)$ is lower and decays with $d_e$. Such effect mitigates with increasing $L$, and the value of $N(d_e+1)/N(d_e)$ should not depend on the exact value of $d_e$ other than its parity in the $L\rightarrow\infty$ limit. The converged values of $N(d_e+1)/N(d_e)$, as shown in Fig. S6, is clearly higher than the reference values, indicating that there exists a positive correlation in errors between adjacent gates. In other word, given that an error occurs at gate $i$, the probability of observing another error at its neighboring gate is enhanced due to the spatial correlation.  

\section{Parity computing device}
\subsection{Computation protocol}
For concreteness, we consider the input sequence $\textbf{X}=\{0,0,1,1,0,1,0,0,1,0,0,0\}$ and plot in Fig. S10 $B$-$C$ the $V_{\textrm{out}}$ of the XOR gates, D1 and D2, for an ensemble of trajectories. Here the clock cycle is chosen as $\tau_c=10^7\beta\hbar$. All gates are operated at $V_d=8V_T$, while all capacitors are initialized with zero charge. We highlight with red cross the time points where outputs are being read out from the D flip-flops. At $t=0$, we send in 4 input data, $X_1$ to $X_4$, by connecting all the input two-way switches to terminal 1. Since all the D flip-flops are slack at the moment, we can store the computing results of the XOR gates into D1 and D2 by switching both output two-way switches to terminal 1 as well. The clock stays at 0 within the first half cycle while computations are being done at the ALU, until $t=\tau_c/2$, when the clock switches to 1 and the outputs of the ALU are being written into D1 and D2. At $t=\tau_c$, as the clock returns to 0, another 4 input data are sent in while the output two-way switches are connected to terminal 2 so that outputs can be sent to and stored at D3 and D4. At the end of the second cycle, when we realize that our memory devices are full and can not take in new inputs, we read out the outputs at D1 to D4 and send them back to the ALU as inputs by connecting all input two-way switches to 2. We continue the computation in this manner, until all input data are taken into account and the final output is read from D1 at $t=6\tau_c$.

\subsection{Error analysis in the parity computing device}
The computational tree graph of the computing device is shown in Fig. S10$A$, with 12 input nodes in the zeroth layer representing the input data, and a single final output $l_4^{(1)}$ that computes the parity of the inputs. The nodes in layers 1 to 3 represent intermediate computation results. If an error is observed in any of the nodes whose layer is deeper than 0, we look further at its parent nodes to trace where the error originates, and its child node (if existing) to see how far the error propagates. We call such a record of error vertically along the computational tree graph an error path, and its length is denoted as $l_e$. In this computational tree graph, the maximum value for $l_e$ is 4, which means the error propagates from layer 1 all the way to the final output; while the minimum value for $l_e$ is 1. Among the $1.28\times 10^4$ simulations we have done with different input sequences, we make a histogram of the error paths with length $l_e$ for different clock cycle $\tau_c$, which is shown in Fig. S10$D$. For the shortest clock cycle plotted $\tau_c=5\times 10^6\beta\hbar$, which is too soon for the gates to reach their steady states, we observe an overwhelmingly high number of error paths of length $l_e=4$. However, when $\tau_c$ is longer, we see an exponential decay in the number of error paths with increasing $l_e$. This exponential decay rate characterizes the temporal correlation between intermediate computation results. The rate increases with longer $\tau_c$, indicating the diminishing correlations, or the weakening of the memory effect at longer clock cycle. With $\tau_c>2\times 10^7\beta\hbar$, it is almost impossible to find an error path with $l_e=4$ that propagates through the computational tree graph.

\begin{figure}
\centering
\includegraphics[width=\textwidth]{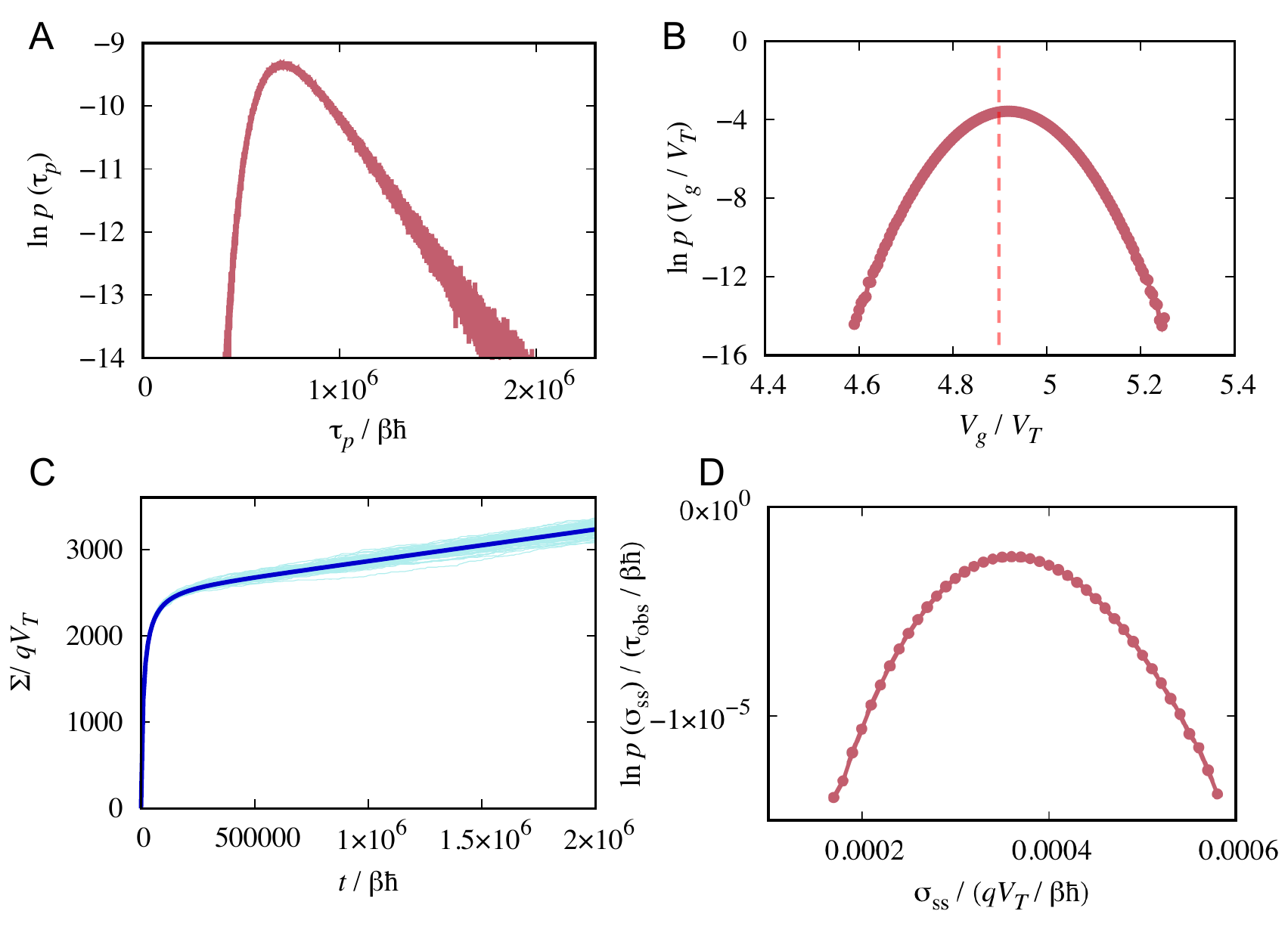}
\caption{Dynamics of a single NOT gate with $X=0$, $V_d=5V_T$, and $V_g$ initialized from zero. (\textit{A}) Distribution of the propagation delay $\tau_p$. (\textit{B}) Distribution of $V_g$ at $t=5\times 10^6\beta\hbar$, where the red dashed line labels the threshold voltage for $Y=1$. (\textit{C}) Evolution of the total heat dissipation of 100 individual simulated trajectories (light blue) and their ensemble average (dark blue). (\textit{D}) Distribution of the entropy production rate in the steady state, measured in the long time limit where $\tau_{\textrm{obs}}=10^6\beta\hbar$. }
\end{figure}

\begin{figure}
\centering
\includegraphics[width=0.6\textwidth]{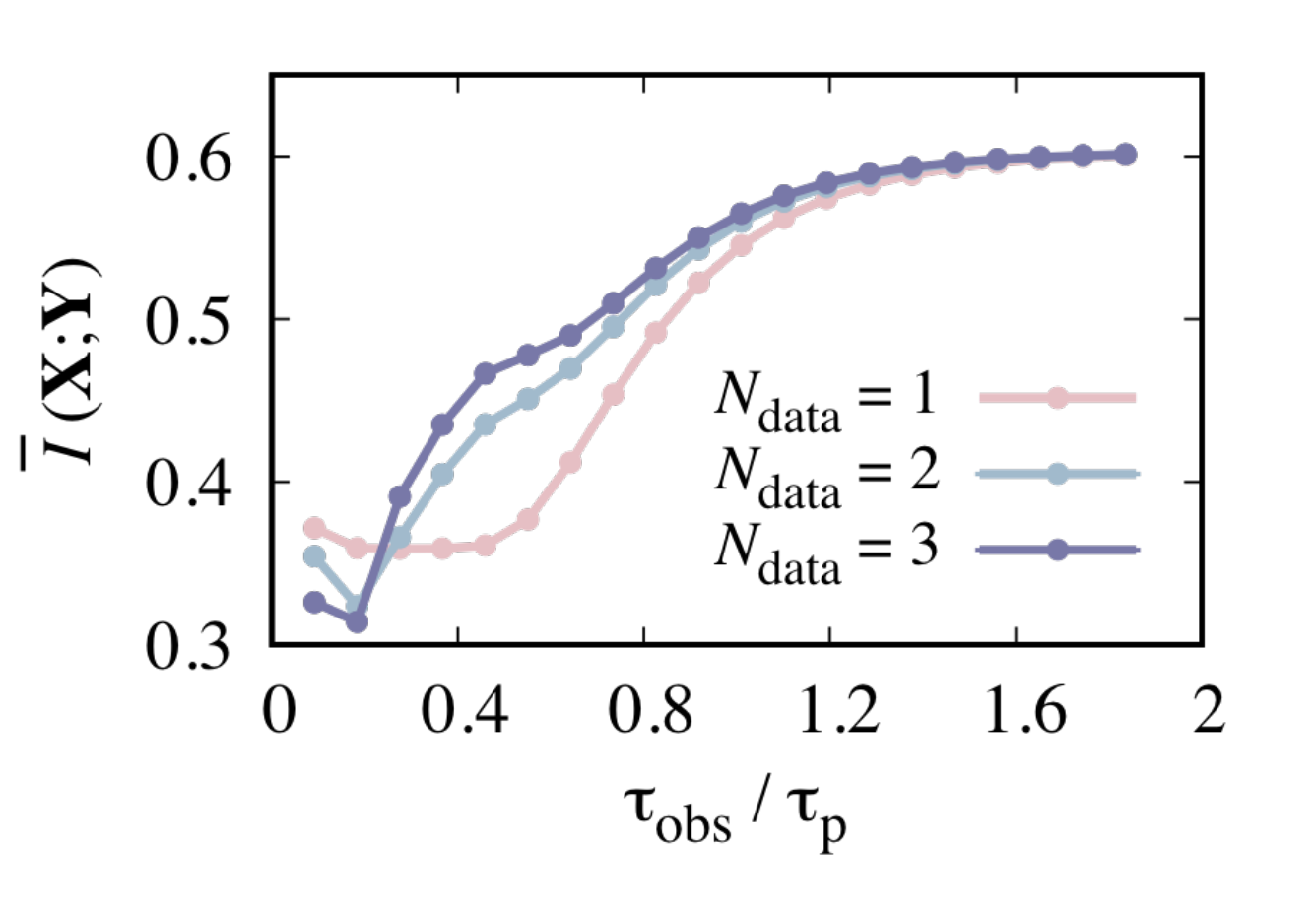}
\caption{Information rate for a repeatedly used NOT gate with $V_d=5V_T$ as a function of the data processing time $\tau_{\textrm{obs}}$.}
\end{figure}

\begin{figure}
\centering
\includegraphics[width=\textwidth]{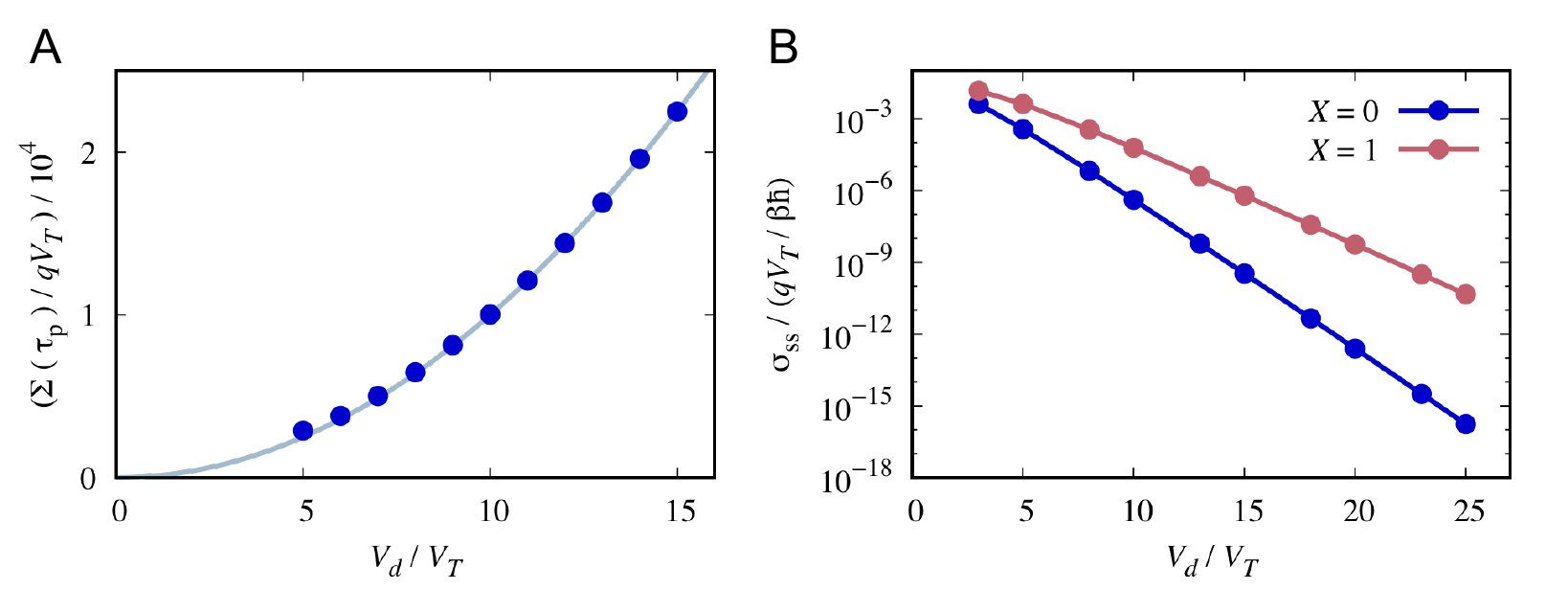}
\caption{Heat dissipation as a function of $V_d$ for a NOT gate with $V_g$ initialized to $0V_T$. (\textit{A}) The heat dissipation during the propagation delay $\tau_p$ (dots) for $X=0$, which converges to the quadratic function $C_gV_d^2/2$ (line) at high $V_d$. (\textit{B}) The heat dissipation rate in the steady state $\sigma_{\textrm{ss}}$ decreases exponentially with $V_d$ for both $X=0/1$.   }
\end{figure}

\begin{figure}
\centering
\includegraphics[width=0.6\textwidth]{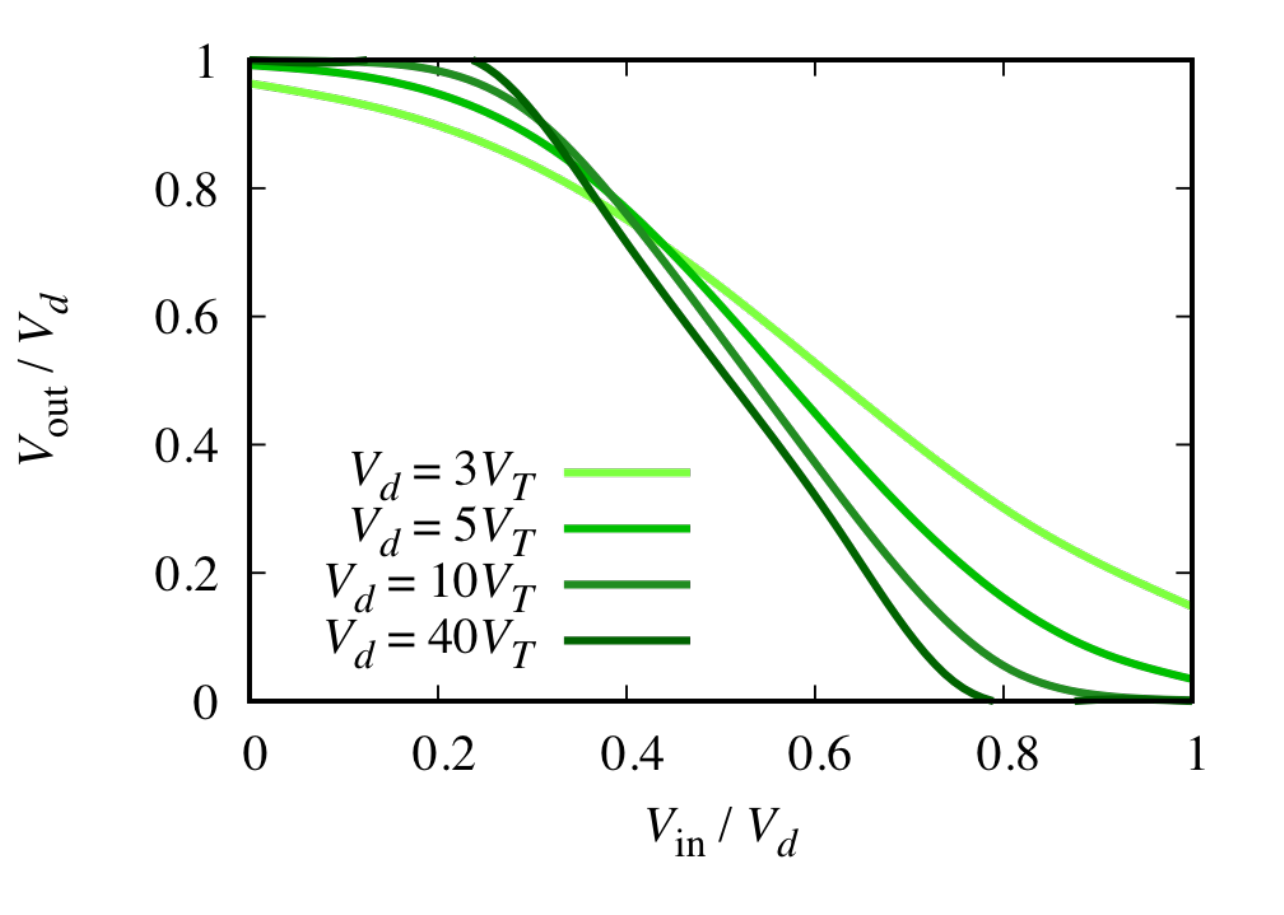}
\caption{The voltage transfer curve of the NAND gate along the one dimension $V_{\textrm{in}}=V_{\textrm{in,A}}=V_{\textrm{in,B}}$, which becomes more symmetric as $V_d$ increases.}
\end{figure}

\begin{figure}
\centering
\includegraphics[width=\textwidth]{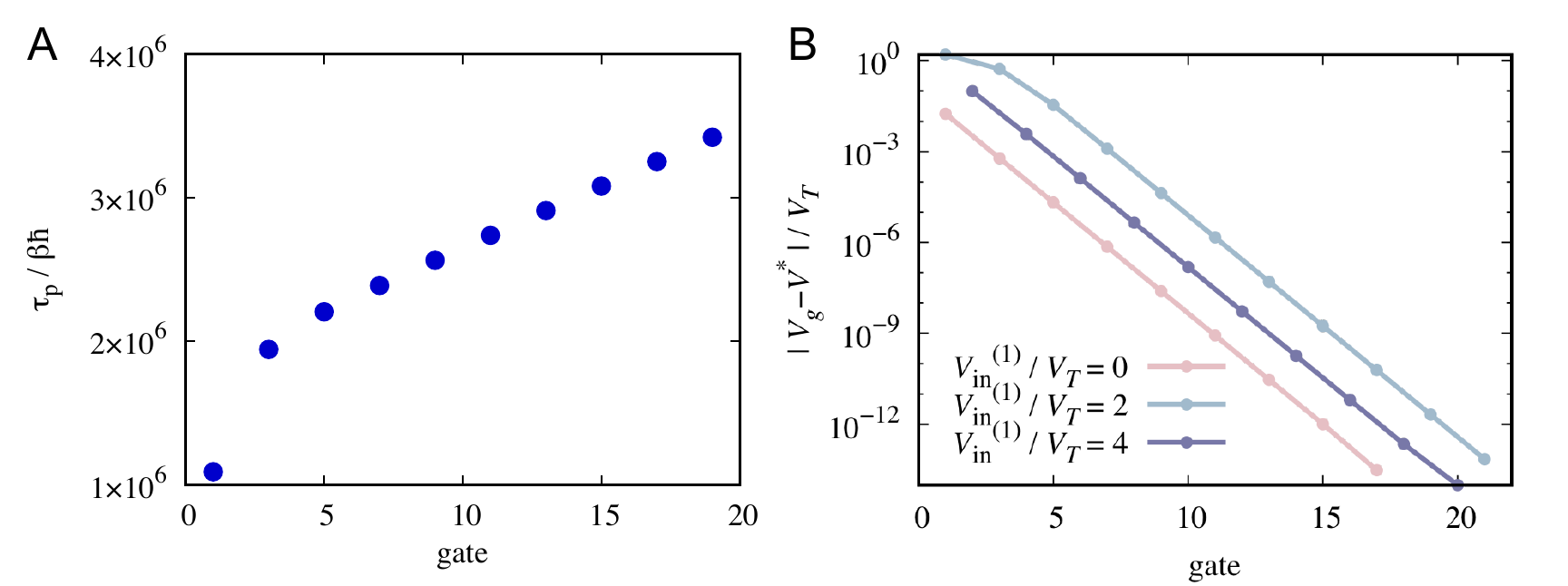}
\caption{Spatial propagation in an array of NOT gates with $V_d=5V_T$ and all $V_g$ initialized to $0V_T$. (\textit{A}) Propagation delay for the odd gates with $X=0$. (\textit{B}) Exponential decay of the distance between $V_g$ and the fixed point. The spatial propagation rate $\kappa$ is deduced from the rate of the exponential decay, which is independent of the input $V_{\textrm{in}}^{(1)}$.}
\end{figure}

\begin{figure}
\centering
\includegraphics[width=0.6\textwidth]{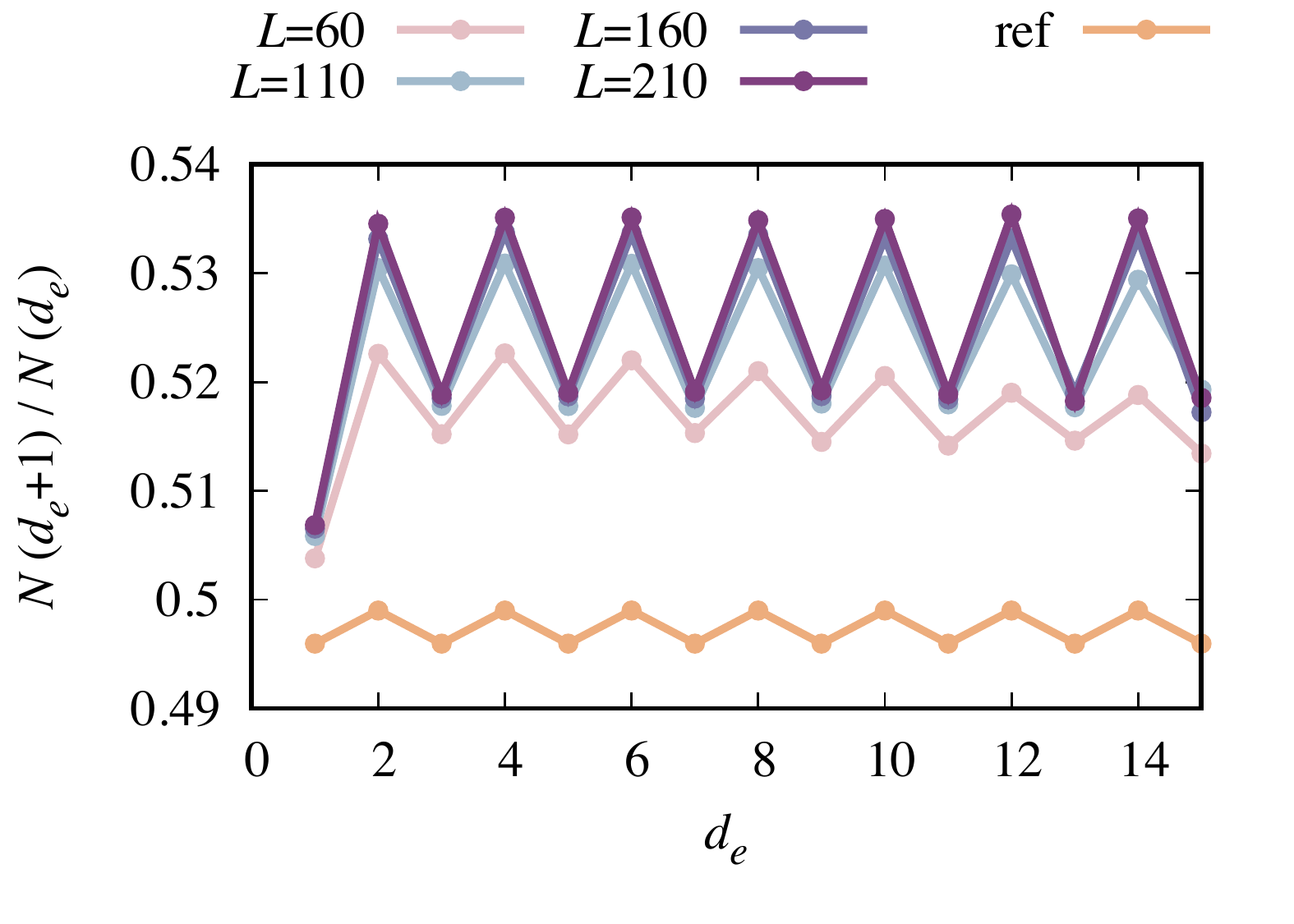}
\caption{The ratio $N(d_e+1)/N(d_e)$ as a function of $d_e$ calculated for an array of NOT gates of length $L=60,110,160$ and 210 with $V_d=5V_T$. The orange dots represent the reference where all gates are independent from each other, and alternates between $2\xi_0\xi_1/(\xi_0+\xi_1)$ and $(\xi_0+\xi_1)/2$.}
\end{figure}

\begin{figure}
\centering
\includegraphics[width=0.6\textwidth]{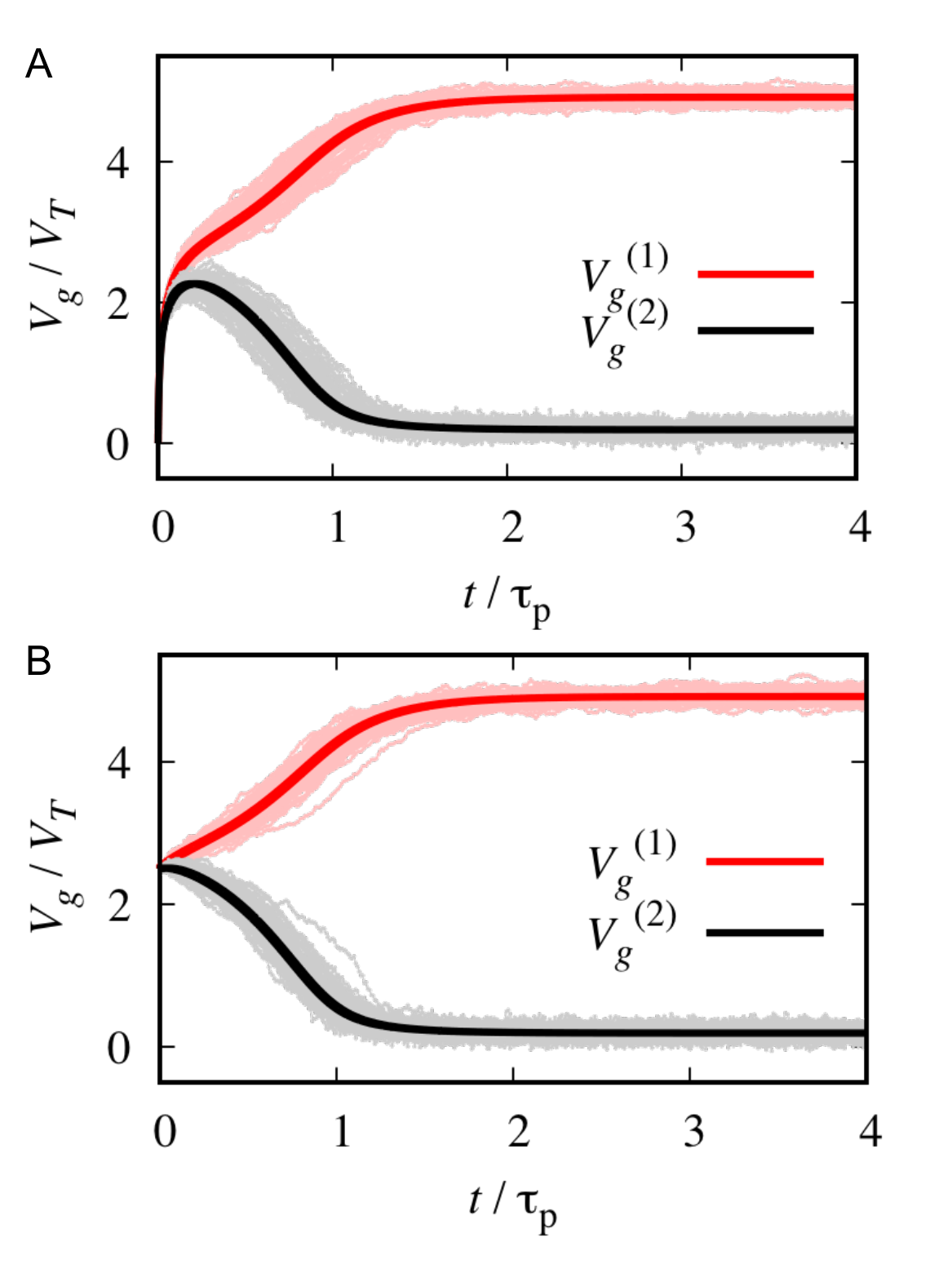}
\caption{The evolution of the outputs $V_g^{(1)}$  and $V_g^{(2)}$ with the initialization $V_g^{(1)}=V_g^{(2)}=0V_T$ (\textit{A}) and $V_g^{(1)}=V_g^{(2)}=2.5V_T$ (\textit{B}) where $V_d=5V_T$. The dark curves represent the steady state solution from the average protocol, while the lighter curves represent 100 trajectories. In both cases, the percentage of trajectories that end up in the other informational state $(V_g^{(1)},V_g^{(2)})=(0.20V_T, 4.89V_T)$ is less than $10^{-5}$ among the $2.56\times 10^6$ trajectories simulated.}
\end{figure}

\begin{figure}
\centering
\includegraphics[width=0.5\textwidth]{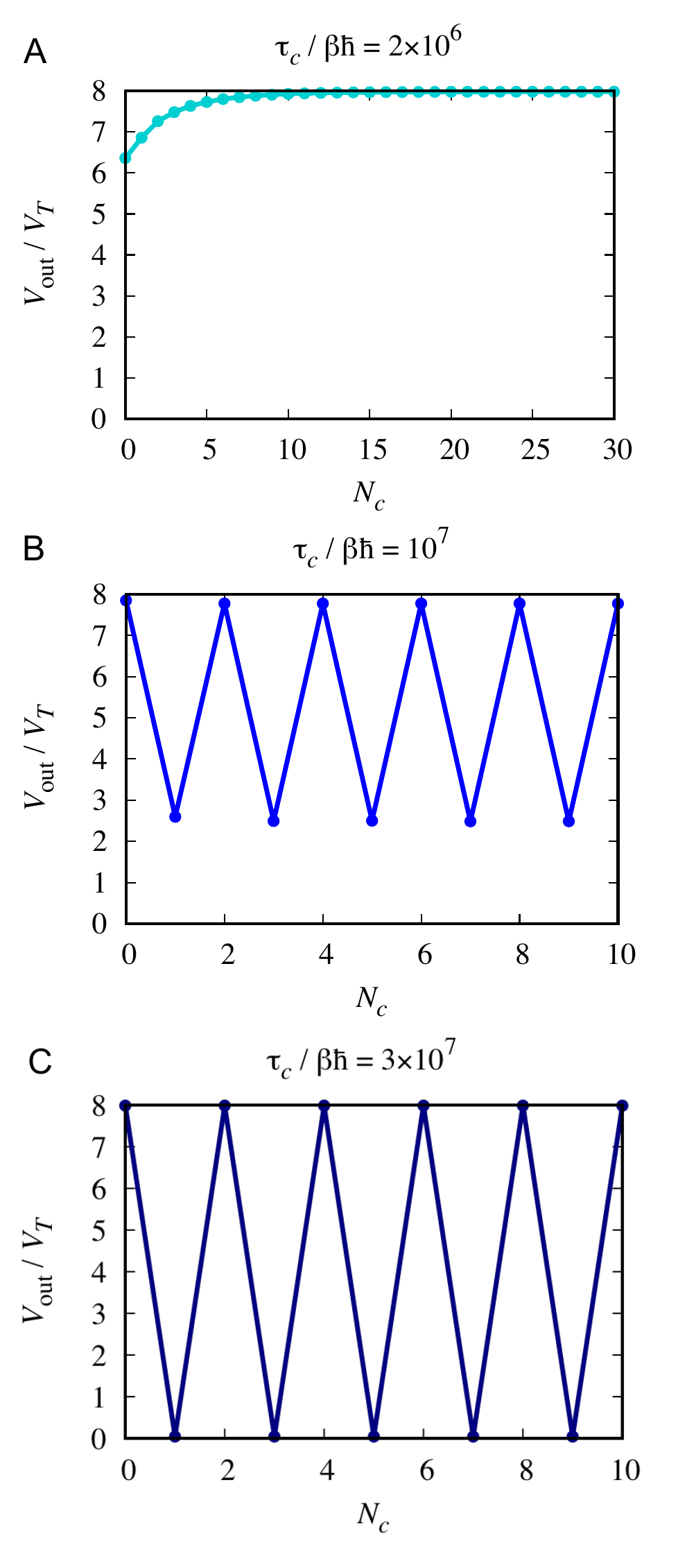}
\caption{Output voltage of a Dflipflop with clock cycle $\tau_c=2\times 10^6\beta\hbar$ (\textit{A}), $\tau_c= 10^7\beta\hbar$ (\textit{B}), and $\tau_c=3\times 10^7\beta\hbar$ (\textit{C}). The input data sequence starts from $X_{\textrm{data}}=1$ and alternates between 1 and 0. All gates are operated at $V_d=8V_T$, and are initialized with $V_g=0V_T$. We discard the first few cycles and average over more than 50 cycles when computing the average error rate and dissipation in Fig. 6, 
so that their values have no dependence on the initialization. }
\end{figure}

\begin{figure}
\centering
\includegraphics[width=0.7\textwidth]{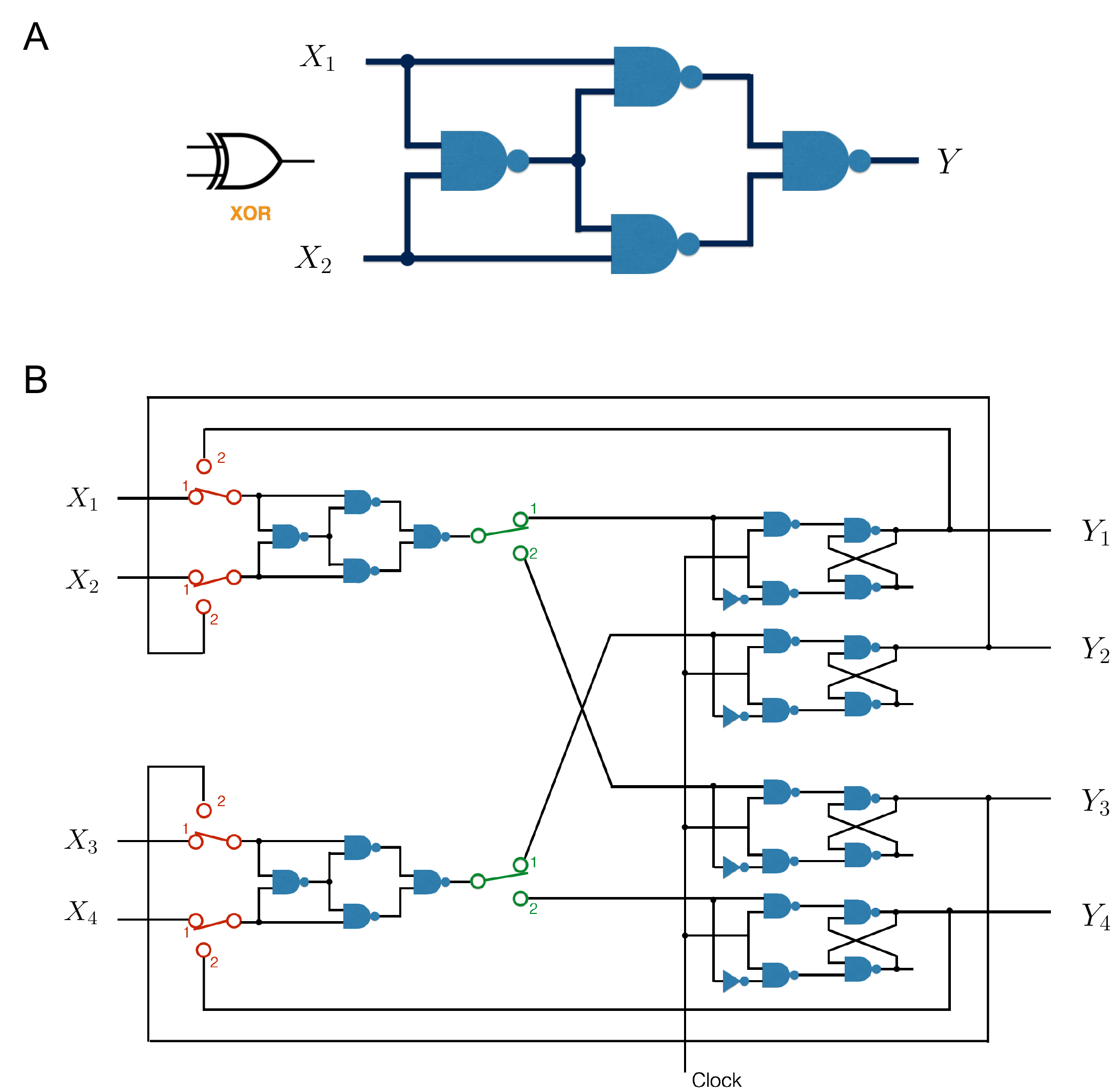}
\caption{(\textit{A}) Symbol (left) and circuit diagram (right) of a XOR gate, which takes in two inputs $X_1$, $X_2$, and compute their parity as output $Y$. (\textit{B}) Circuit diagram of the parity computing device with 2 XOR gates and 4 D flip-flops.}
\end{figure}

\begin{figure}
\centering
\includegraphics[width=0.6\textwidth]{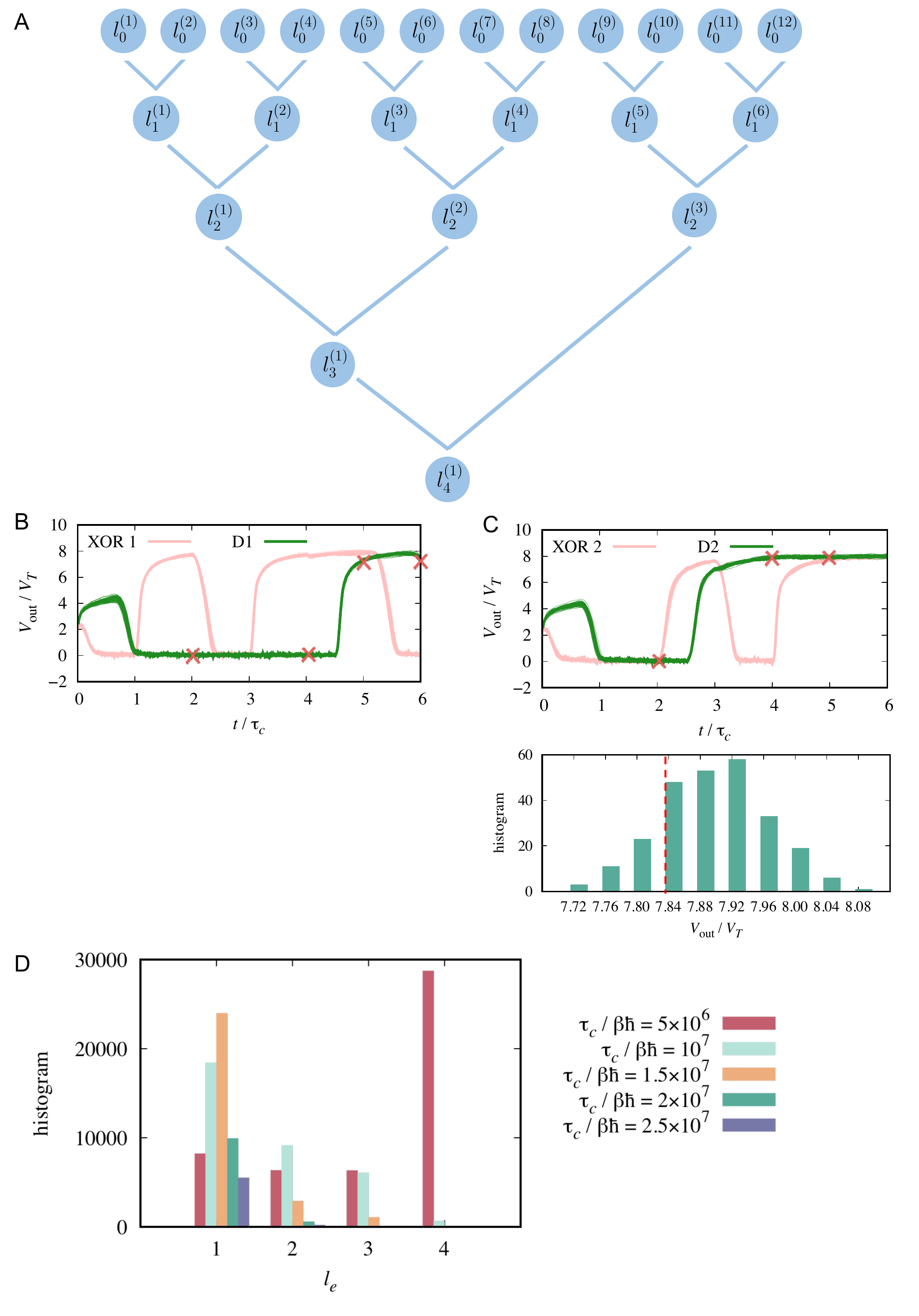}
\caption{(\textit{A}) Computational tree graph of the device that computes the parity of 12 input data. The node $l_i^{(j)}$ denotes the $j$th output of the $i$th layer. The $0$th layer has 12 nodes, which represent the 12 data in the input sequence. Each child node calculates the parity of its two parent nodes. The final output $l_4^{(1)}$ computes the parity of all the input data.
(\textit{B}) For the input sequence $\textbf{X}=\{0,0,1,1,0,1,0,0,1,0,0,0\}$, the output of XOR 1 and its corresponding memory storage D1 for 64 individual trajectories at $V_d=8V_T$, $\tau_c=10^7\beta\hbar$. The red crosses label points where outputs on D1 are read out for further processing. (\textit{C}) The output of XOR 2 and its corresponding memory storage D2 with the same parameters as in \textit{B}. The lower plot shows the histogram of outputs at D2 at $t=4\tau_c$. The red dotted line labels the threshold under which the output corresponds to an error. (\textit{D}) Histogram of error paths of length $l_e$ for different clock cycle $\tau_c$.}
\end{figure}